\documentclass[fleqn,usenatbib]{mnras}

\usepackage[T1]{fontenc}

\DeclareRobustCommand{\VAN}[3]{#2}
\let\VANthebibliography\thebibliography
\def\thebibliography{\DeclareRobustCommand{\VAN}[3]{##3}\VANthebibliography}

\usepackage{graphicx}	
\usepackage{amsmath}	
\usepackage{amssymb}	

\usepackage{newtxtext,newtxmath}

\newcommand{\mdot}{$\mathrm{M}_\odot$}
\newcommand{\zdot}{$\mathrm{Z}_\odot$}
\newcommand{\sphgal}{\textsc{sphgal}}
\newcommand{\gadget}{\textsc{gadget-3}}
\newcommand{\healpix}{\textsc{healpix}}
\newcommand{\treecol}{\textsc{treecol}}
\newcommand{\bifrost}{\textsc{bifrost}}
\newcommand{\ketju}{\textsc{ketju}}
\newcommand{\mstar}{\textsc{mstar}}
\newcommand{\griffin}{\textsc{griffin}}

\title[Star cluster evolution in dwarf galaxies]{The formation, evolution and disruption of star clusters with improved gravitational dynamics in simulated dwarf galaxies}

\author[Lah\'en et al.]
{Natalia Lah\'en$^{1}$\thanks{E-mail: nlahen@mpa-garching.mpg.de},
Antti Rantala$^{1}$,
Thorsten Naab$^{1}$,
Christian Partmann$^{1,2}$,
Peter H. Johansson$^{3}$ and 
\newauthor
Jessica May Hislop$^{4}$ 
\vspace{1mm}
\\
$^{1}$Max-Planck-Institute f\"{u}r Astrophysik, Karl-Schwarzschild-Stra{\ss}e 1, D-85740 Garching, Germany\\
$^{2}$Center for Computational Astrophysics, Flatiron Institute, 162 5th Avenue, New York, NY 10010, US\\
$^{3}$Department of Physics, University of Helsinki, Gustaf Hällströmin katu 2, FI-00014, Helsinki, Finland\\
$^{4}$The Institute of Cancer Research, 123 Old Brompton Road, London SW7 3RP, UK
}

\date{Accepted 2025 February 24. Received 2025 January 13; in original form 2024 October 1}

\pubyear{2025}

\begin{document}
\label{firstpage}
\pagerange{\pageref{firstpage}--\pageref{lastpage}}
\maketitle

\begin{abstract}

So far, even the highest resolution galaxy formation simulations with gravitational softening have failed to reproduce realistic life cycles of star clusters. We present the first star-by-star galaxy models of star cluster formation to account for hydrodynamics, star formation, stellar evolution and collisional gravitational interactions between stars and compact remnants using the updated \textsc{sphgal}+\textsc{ketju} code, part of the \textsc{griffin}-project. Gravitational dynamics in the vicinity of $>3$ M$_\odot$ stars and their remnants are solved with a regularised integrator (\textsc{ketju}) without gravitational softening. Comparisons of idealised star cluster evolution with \textsc{sphgal}+\textsc{ketju} and direct N-body show broad agreement and the failure of simulations that use gravitational softening. In the hydrodynamical simulations of idealised dwarf galaxies run with \textsc{sphgal}+\textsc{ketju}, clusters up to $\sim900$ M$_\odot$ form compact (effective radii 0.1--1 pc) and their sizes increase by up to a factor of ten in agreement with previous N-body simulations and the observed sizes of exposed star clusters. The sizes increase rapidly once the clusters become exposed due to photoionising radiation. On average 63\% of the gravitationally bound clusters disrupt during the first 100 Myr of evolution in the galactic tidal field. The addition of collisional dynamics reduces the fraction of supernovae in bound clusters by a factor of $\sim 1.7$, however the global star formation and outflow histories change by less than 30\%. We demonstrate that the accurate treatment of gravitational encounters with massive stars enables more realistic star cluster life cycles from the earliest stages of cluster formation until disruption in simulated low-mass galaxies.

\end{abstract}




\section{Introduction}

Star clusters are collisional systems whose internal evolution, after gas-removal, is governed by stellar evolution and gravitational interactions between individual stars, their compact remnants and the galactic tidal field. The picture is especially complex at the early stages of cluster evolution due to ongoing star formation and the interactions between young (massive) stars and their gaseous surroundings. Observationally the detailed structure of star clusters within the Local Group can be resolved down to individual stars \citep{2013A&A...558A..53K, 2016MNRAS.458..624C}. Multi-wavelength studies of integrated light in young star clusters have been used to characterise the ages, masses, metallicities, and sizes of young clusters both in the local Universe \citep{1999AJ....118.1551W, 2003AJ....126.1836H, 2017ApJ...841..131A, 2020MNRAS.499.3267A, 2023MNRAS.519.3749C} and at increasingly high redshifts \citep{2024Natur.636..332M, 2024Natur.632..513A} thanks to the superb resolving capabilities of e.g. the \textit{Hubble Space Telescope} (HST), \textit{Very Large Telescope} (VLT) and the \textit{James Webb Space Telescope} (JWST).

Young star clusters have mass functions $dN/dM\propto M^\alpha$ with a power-law index close to $\alpha=-2$ across different star-forming environments \citep{1996ApJ...471..816E,1996ApJ...466..802E,1999ApJ...527L..81Z,2003AJ....126.1836H, 2012ApJ...752...96F, 2019ApJ...872...93M}. A truncation in the mass function at high cluster masses has been suggested \citep{2015MNRAS.452..246A,2018MNRAS.477.1683M, 2020MNRAS.499.3267A}, but quantifying this cut-off mass is difficult due to low-number statistics. The mass-size relation of star clusters is observed to be shallow (half-mass or half-light radii $\propto M^\beta$ with $\beta\lesssim 1/3$, see the data collected in \citealt{2019ARA&A..57..227K} and \citealt{2021MNRAS.508.5935B}). These measurements exclude the youngest, embedded phase of early cluster evolution that is only revealed at rest-frame infrared wavelengths. The unprecedented sensitivity of JWST enables the detection of embedded star formation in the form of individual proto-stars down to sub-solar masses within the nearby galactic environments \citep{2017ApJ...846..110D, 2023NatAs...7..694J}, thus clear improvements to the measurements of embedded cluster sizes can be expected in near future. 

Both the amount of star formation occurring in bound clusters and the amount of disruption that clusters undergo are still uncertain. The cluster formation efficiency (CFE or $\Gamma$, \citealt{2008MNRAS.390..759B}) defines the fraction of stellar mass forming in bound clusters. It has been studied within galaxies \citep{2013MNRAS.436L..69S, 2016ApJ...827...33J, 2018MNRAS.477.1683M} and across galaxy samples \citep{2010MNRAS.405..857G, 2020MNRAS.499.3267A, 2023ApJ...949..116C, 2023MNRAS.519.3749C} but its correlation with the star-forming environment has not been firmly established (see the discussion e.g. in \citealt{2017ApJ...849..128C} and \citealt{2020SSRv..216...69A}). Analytic models \citep{2012MNRAS.426.3008K} and numerical star formation simulations \citep{2019MNRAS.490.1714P, 2022MNRAS.514..265L, 2023MNRAS.519.1366G} often find that $\Gamma$ increases with increasing star formation rate (SFR) surface density in massive or starbursting galaxies while the scatter in $\Gamma$ increases toward lower galaxy mass and no environmental correlation is found \citep{2022MNRAS.509.5938H, 2023MNRAS.522.3092L, 2024A&A...681A..28A}. Some observations indicate that the cluster disruption rate depends on the cluster mass \citep{2005A&A...441..117L, 2012MNRAS.419.2606B} while others find a universal disruption rate \citep{2010ApJ...711.1263C, 2012ApJ...752...96F}. The disruption rate has been suggested to depend on the galactic environment \citep{2012MNRAS.419.2606B, 2018MNRAS.477.1683M} with clusters in galactic outskirts or low-mass galaxies disrupting on a longer time-scale \citep{2024A&A...687A.101A}. Numerical work has shown that the response of the cluster to gas-removal and the overall mass-loss can depend on the initial structure of the cluster with respect to the tidal field \citep{2003MNRAS.340..227B, 2007MNRAS.380.1589B,  2010MNRAS.409..305L, 2013MNRAS.428.1303S, 2018ApJ...863..171S}. Quantifying the initial sizes and very early evolution of star clusters is thus important for understanding their long-term evolution.

The distribution of gas around young clusters has been used to estimate the time-scale of gas expulsion \citep{2011ApJ...729...78W, 2017A&A...601A.146C, 2018MNRAS.481.1016G, 2021ApJ...909..121M, 2022MNRAS.512.1294H}, in order to quantify the role of various stellar feedback processes in determining the initial cluster properties.  Recently exposed clusters are often very young ($<5$ Myr) and have been associated with gaseous outflows \citep{2021ApJ...912....4L, 2024AJ....167..166S} conceptually similar to expanding superbubbles \citep{2023A&A...676A..67W}. With the exposed clusters extending to ages as low as \mbox{$\sim2$ Myr}, pre-supernova feedback (HII regions and stellar winds) has been proposed as an important driver of gas expulsion in young star clusters even before the destructive supernovae (SNe) occur. Counter-examples exist as well: \citet{2023ApJ...944L..20K} and \citet{2023ApJ...946....1C} find massive, embedded clusters that are 5--6 Myr old, indicating that massive and/or compact enough clusters can resist gas expulsion. The majority of intermediate age and old massive globular clusters (GCs) exhibit chemical variations but only in their light-element abundances \citep{2018ARA&A..56...83B, 2019A&ARv..27....8G}. This could be the smoking gun of gas retention in massive clusters that undergo star formation for an extended period of time while massive stars release chemically enriched stellar winds \citep{2016A&A...587A..53K, 2019ApJ...871...20S, 2024MNRAS.530..645L}. What role cluster mass, compactness and metallicity play in the ability of a star cluster to continue forming stars even after the ignition of the first massive stars still remain open questions. \citet{2019MNRAS.489.1880H} have for instance found in molecular cloud simulations that regardless of the initial mass or the compactness of the progenitor gas clouds, they will continue forming stars for a few sound crossing times after which radiation disperses the remaining gas. However, galaxy scale simulations are needed to address the full life cycle of molecular clouds and star clusters (see e.g. the discussion in \citealt{2024MNRAS.527.7093J}).

Resolving the initial properties of star clusters and the dominant sources of feedback energy on scales of individual stars is difficult in observations of clustered environments. High-resolution hydrodynamical simulations of galactic star formation can shed light on these complex processes that originate from small spatial scales \citep[see e.g.][]{2017ARA&A..55...59N}. Individual (massive) stars and their stellar feedback processes have been included in galaxy-scale simulations to study e.g. galactic interstellar medium (ISM) structure and outflows \citep{2017MNRAS.471.2151H, 2020MNRAS.495.1035S, 2021MNRAS.501.5597G, 2024ApJ...960..100S}, star formation efficiency \citep{2022MNRAS.509.5938H}, cosmological star formation history and Population III stellar feedback in low-mass galaxies \citep{2022MNRAS.513.1372G, 2024ApJ...970...14S, 2025ApJ...978..129A}, clustering of stars and their feedback in idealised \citep{2020ApJ...891....2L, 2021MNRAS.506.3882S, 2024A&A...691A.231D, 2024MNRAS.530..645L} and cosmological settings \citep{2022MNRAS.516.5914C, 2023MNRAS.522.2495G, 2024ApJ...971..103G, 2024arXiv241102502C}, and the impact of runaway stars on star formation and outflows \citep{2020MNRAS.494.3328A, 2023MNRAS.526.1408S, 2023MNRAS.521.2196A}. \citet{2023MNRAS.522.3092L} introduced a numerical model (based on \citealp{2017MNRAS.471.2151H} and  \citealp{2019MNRAS.483.3363H} and references therein)  within the Galaxy Realizations Including Feedback From INdividual massive stars\footnote{\url{https://www.mpa-garching.mpg.de/~naab/griffin-project}} (\griffin) project which considers every newly formed star as an individual particle. The inclusion of stellar evolution models describing the mass, energy, and metal output of massive and very massive stars throughout their lives enables the exploration of their role in the formation and evolution of star clusters star-by-star in their galactic environment. En route to building a self-consistent picture of how GCs form in a galactic context, \citet{2024MNRAS.530..645L} analysed the early chemical enrichment in proto-GCs and demonstrated that massive clusters can retain gas and recycle stellar wind-material efficiently compared to SN-material. 

However, one key ingredient missing in previous studies of galactic-scale star formation is the collisional treatment of the individually realised stars, which we address in this paper. Gravitational two-body and multi-body interactions drive the internal evolution of star clusters through relaxation, mass-segregation, dynamical binary formation and core collapse \citep{1971ApJ...164..399S, Spitzer1987, Goodman1993}, all of which are suppressed when gravitational softening is used to smooth out the gravitational potential in close encounters. Concurrent modelling of star formation, stellar feedback and collisional N-body dynamics has been successfully demonstrated in simulations of individual giant molecular clouds and star-forming regions \citep{2020ApJ...904..192W, 2020MNRAS.499..748D, 2021PASJ...73.1057F, 2021MNRAS.506.2199G, 2022MNRAS.509.6155R}. 

Collisional simulation codes typically use higher-order integration methods such as the fourth-order Hermite \citep{1991ApJ...369..200M} or the fourth-order forward integrator \citep{Rantala2021} which provide better integration accuracy at the same computational cost compared to e.g. the standard leap-frog algorithm widely used in galactic-scale codes. These fourth-order algorithms are usually, though not always, coupled to regularised few-body solvers. Hydrodynamical simulation codes can be coupled with dedicated N-body solvers such as the direct summation code \textsc{ph4} \citep{2012ASPC..453..129M} or the hybrid \textsc{PeTaR} code \citep{2020MNRAS.497..536W} implemented in the \textsc{AMUSE} framework \citep{2009NewA...14..369P, 2013A&A...557A..84P}. 

Directly summing all two-body forces in a galaxy-scale simulation would however be prohibitively inefficient. \citet{2024ApJ...974..193J} recently implemented the N-body code \textsc{nbody6++gpu} \citep{2015MNRAS.450.4070W} in the \textsc{enzo} code, which they used to solve the collisional evolution of idealised star clusters in a hydrodynamical Milky Way-mass galaxy. In this study we use the regularised tree code \ketju{} \citep{2017ApJ...840...53R, 2023MNRAS.524.4062M}, which includes a number of numerical algorithms to enable fast and accurate integration of selected simulation regions using the \mstar-integrator \citep{2020MNRAS.492.4131R}. \ketju{} was first developed in \gadget{} to be primarily used in larger scale simulations to model the dynamics of supermassive black holes and supermassive black hole binaries in their stellar-dynamical environments \citep{2017ApJ...840...53R, 2018ApJ...864..113R, 2021ApJ...912L..20M, 2023MNRAS.520.4463L, 2023MNRAS.524.4062M}. 

\citet{2025MNRAS.537..956P} introduced recently \ketju{} in the \sphgal{} version of \gadget{} in a new suite of star-by-star \griffin-simulations where \ketju{} was used to compute accurate interactions of central massive black holes with surrounding individual stars while interactions between stars were still softened. Here we expand the code-implementation of \citet{2025MNRAS.537..956P} by adding the \ketju{} regularised integration regions around all forming massive stars in an entire galaxy. We execute comparison simulations with \sphgal{}-\ketju{} and the direct N-body code \bifrost{} \citep{2023MNRAS.522.5180R, 2024MNRAS.531.3770R}. Star clusters modelled using \ketju{} expand in size and undergo dynamical mass-loss. We then run hydrodynamical simulations with star formation and an evolving galactic tidal field resembling that of the Wolf–Lundmark–Melotte galaxy as described in \citet{2017MNRAS.471.2151H}. An accurate accounting of gravitational encounters with massive stars produces more realistic star clusters. The clusters form compact and expand rapidly once the first massive stars ignite and expel any remaining gas. 

This article is structured as follows. Section \ref{section:simulations} provides an overview of the \sphgal{} simulation method and describes briefly the updated interface with the \ketju-integrator module. We also give a brief introduction of the direct N-body code \bifrost{} used to validate the \sphgal+\ketju{} code. A description of the code comparison tests and galaxy initial conditions is given here as well. Section \ref{section:nbody_results} describes the results of the code comparison, where we investigate the evolution of the size, density and velocity distribution of idealised star clusters run in isolation. Section \ref{section:hyd_results} discusses the formation, evolution, and disruption of star clusters in hydrodynamical simulations of isolated low-metallicity dwarf galaxies run with the updated \sphgal+\ketju{} code, in comparison to simulations that use gravitational softening in all gravitational particle interactions. Our conclusions and final remarks are presented in Section \ref{section:conclusions}.

\section{Simulations}\label{section:simulations}

\subsection{Hydrodynamical simulation code \sphgal}

The simulations for this study were run with the \sphgal-version \citep{2014MNRAS.443.1173H, 2016MNRAS.458.3528H, 2017MNRAS.471.2151H, 2019MNRAS.483.3363H} of the \gadget{} code \citep{2005MNRAS.364.1105S}. We used the modern smoothed particle hydrodynamics (SPH) implementation in \sphgal, which includes several improvements to the numerical accuracy of the SPH methodology as outlined e.g. in \citet{2014MNRAS.443.1173H,2016MNRAS.458.3528H}. Recent upgrades (\citealt{2023MNRAS.522.3092L}) to the code include a method for locally mass-conserving sampling of individual stars from a given stellar initial mass function (IMF) and stellar models that describe the evolution of massive stars described briefly below. Here we expand the methodology for accurate treatment of collisional gravitational dynamics introduced in \citet{2025MNRAS.537..956P} to include high-accuracy \ketju{} \citep{2017ApJ...840...53R} integration regions around all massive stars and their remnants. In the simulations, we assume an initial metallicity of $Z\sim 0.01$ \zdot{} and adopt a fiducial resolution of 4 \mdot{} in gas and old stellar disk particles and $6.8\times10^3$ \mdot{} in dark matter. The stars formed during the simulation have masses sampled from an IMF down to the hydrogen burning limit of 0.08 \mdot{}.

\subsubsection{Cooling, chemistry and star formation}

As outlined in detail in \citet{2016MNRAS.458.3528H} and \citet{2017MNRAS.471.2151H}, cooling and chemistry of the ISM are treated in two temperature regimes. We use a chemical network at low temperatures ($<3\times10^4$ K) and tabulated metal-dependent cooling rates from \citet{2009MNRAS.393...99W} at high temperatures ($>3\times10^4$ K). The chemistry network includes six chemical species (H$_2$, H$^+$, H, CO, C$^+$, O) and free electrons, based on the methods of \citet{1997ApJ...482..796N}, \citet{2007ApJS..169..239G} and \citet{2012MNRAS.421..116G} closely following the implementation in the \textsc{SILCC}-project \citep{2015MNRAS.454..238W}. The chemistry network takes into account the spatially resolved interstellar radiation field that is attenuated based on the gas and dust column density as outlined in Section \ref{section:FB}. The code tracks the abundances of 13 individual elements in gas and stars: H, He, N, C, O, Si, Al, Na, Mg, Fe, S, Ca and Ne.

For star formation, we consider a threshold according to the local Jeans-mass ($M_\mathrm{J}$) estimate within the SPH kernel, using a definition for $M_\mathrm{J}$ as
\begin{equation}
   M_\mathrm{J} = \frac{\pi^{5/2}c_\mathrm{s}^3}{6G^{3/2}\rho^{1/2}}, 
\end{equation}
where $c_\mathrm{s}$ is the sound speed, $\rho$ is the SPH-averaged gas density and $G$ is the gravity constant. When $M_\mathrm{J}$ drops below half of the SPH kernel mass (\mbox{$0.5\times 400$ \mdot}), gas particles are turned into star particles. The star particles are first considered to be \textit{reservoir} particles for a dynamical time according to the local gas density, $t_\mathrm{dyn}=(4\pi G \rho)^{-1/2}$. This approximates further gravitational collapse of the parental dense gas phase. In this phase the stellar reservoir particles are decoupled from the hydrodynamics and interact via gravitational forces only. After one dynamical time, the reservoir particle mass is sampled into individual stars along the \citet{2001MNRAS.322..231K} IMF between 0.08 \mdot{} and 500 \mdot. To conserve mass on scales comparable to individual star-forming regions, we perform the IMF-sampling particle by particle considering only the combined mass of the reservoir particles within the Jeans-length $R_\mathrm{J}$ that was measured at conversion, defined as
\begin{equation}
   R_\mathrm{J} = \left( \frac{3}{4\pi} \frac{M_\mathrm{J}}{\rho} \right) ^{1/3}.
\end{equation}
In other words, when the stellar mass sampled per particle exceeds the fiducial resolution of 4 \mdot, the overshoot mass is borrowed from other nearby reservoir particles within $R_\mathrm{J}$. If there is not enough reservoir particle-mass nearby, the last stellar mass is re-sampled. This results in under-sampling of massive stars in regions where the local reservoir for sampling is small (low-SFR regions), while regions of intense star formation (e.g. in a starburst) will result in more fully populated input IMFs. 

\begin{table*}
\begin{tabular}{l c c c c c c c}
\hline
Name & $N_*$ & $N_\mathrm{sample}$ & Time & Code & $\epsilon_*$ & $r_\mathrm{\ketju}$ [pc] & \ketju{} $m_i$ [\mdot]\\
\hline
\texttt{nb\_1k\_BIFROST} & 1155 & 10 & 50 Myr & \bifrost & - & - & -\\
\texttt{nb\_1k\_KETJU\_m0\_e001} & 1155 & 10 & 50 Myr & \sphgal+\ketju & $0.01$& $0.03$ & $0.08$\\
\texttt{nb\_1k\_KETJU\_m3\_e001} & 1155 & 10 & 50 Myr & \sphgal+\ketju & $0.01$& $0.03$ & $3$ \\
\texttt{nb\_1k\_KETJU\_m3\_e01} & 1155 & 10 & 50 Myr & \sphgal+\ketju & $0.1$& $0.3$ & $3$ \\
\texttt{nb\_1k\_KETJU\_m8\_e001} & 1155 & 10 & 50 Myr & \sphgal+\ketju & $0.01$& $0.03$ & $8$ \\
\texttt{nb\_1k\_GADGET\_e001} & 1155 & 10 & 50 Myr & \sphgal & $0.01$& - & -\\
\hline
\texttt{nb\_10k\_BIFROST} & 10660 & 10 & 50 Myr & \bifrost & - & - & -\\
\texttt{nb\_10k\_KETJU\_m3\_e001} & 10660 & 10 & 50 Myr & \sphgal+\ketju & $0.01$ & $0.03$ & $3$ \\
\texttt{nb\_10k\_KETJU\_m8\_e001} & 10660 & 10 & 50 Myr & \sphgal+\ketju & $0.01$ & $0.03$ & $8$ \\
\texttt{nb\_10k\_SPHGAL\_e001} & 10660 & 10 & 50 Myr & \sphgal & $0.01$ & - & -\\
\hline
\texttt{hyd\_KETJU\_m3\_e01} & 513794 & 1 & 500 Myr & \sphgal+\ketju & $0.1$ & $0.3$ & $3$ \\
\texttt{hyd\_SPHGAL\_e01} & 489226 & 1 & 500 Myr & \sphgal & $0.1$ & - & -\\
\texttt{hyd\_KETJU\_m3\_e001} & 478024 & 1 & 400 Myr & \sphgal+\ketju & $0.01$ & $0.03$ & $3$ \\
\texttt{hyd\_SPHGAL\_e001} & 487158 & 1 & 500 Myr & \sphgal & $0.01$ & - & -\\
\hline
\end{tabular}
\caption{Description of the N-body simulations. The columns give the simulation name, the number of stars subject to \ketju-integration, the number of random initialisations, the total simulation time, the code used, the stellar gravitational softening length, the size of the regularised region and the initial stellar mass of particles used as centres of \ketju-regions. Run \texttt{hyd\_KETJU\_m3\_e001} was limited to a total runtime of 400 Myr due to the increased computational cost.}
\label{table:ics}
\end{table*}

\subsubsection{Stellar feedback}\label{section:FB}

The individually sampled stars in the simulation release radiation, energy, momentum and chemically enriched matter according to their initial stellar masses, metallicities, and evolutionary stages. All stars are assigned with fixed mass-dependent far-ultraviolet luminosities integrated in the range \mbox{$6$--$13.6$ eV}. For stars below \mbox{9 \mdot} we combine the \textsc{BaSeL} spectral library at $Z\sim 0.01$ \zdot{} \citep{1997A&AS..125..229L, 1998A&AS..130...65L, 2002A&A...381..524W} and the \textsc{geneva} stellar models at $Z=0.0004\sim 0.02$ \zdot \citep{2019A&A...627A..24G}\footnote{Stars below the lower mass limit of \mbox{1.7 \mdot} in the $Z=0.0004$\mbox{$\sim 0.02$ \zdot{}} Geneva tables are supplemented with fluxes from the $Z=0.002$\mbox{$\sim 0.1$ \zdot{}} tables of \citet{2013A&A...558A.103G} between \mbox{$0.8$--$1.7$ \mdot{}} by scaling up the mass-dependent fluxes by a factor of 2, which results in approximately continuous mass-flux relation across the mass-range. Below \mbox{$0.8$ \mdot{}} the fluxes are extrapolated as $L\propto M^{3.5}$.}. The resulting interstellar radiation field within the galaxy, produced by the stellar distribution, is attenuated at the location of each gas particle accounting for dust shielding and gas self-shielding, assuming optically thin conditions. We use the \treecol-algorithm \citep{2012MNRAS.420..745C} to compute the attenuation around a gas particle, in 12 solid-angles divided equally using the \healpix-algorithm \citep{2011ascl.soft07018G}. 

Stars more massive than \mbox{$9$ \mdot{}} are supplemented with radiation and stellar wind properties that evolve according to the stellar age. We adopt the Bonn Optimized Stellar Tracks (BoOST,  \citealt{2022A&A...658A.125S}) that describe the evolution of stars between 9--500 \mdot. We track the far-ultraviolet and photoionising ($>13.6$ eV) luminosities, and the stellar winds characterised by the mass-loss rate, velocity and chemical composition. Photoionisation is implemented using the Str\"omgren approximation by setting the gas to be fully ionised at \mbox{$10^4$ K} within the Str\"omgren sphere of any photoionising star, iteratively in case overlapping HII regions occur (see Section 2.5.2 in \citealt{2017MNRAS.471.2151H}). Stellar winds are released by injecting momentum and metal-enriched material into $12\times 8(\pm2)$ gas particles in 12 equally divided solid angles around the star (see details in \citealt{2023MNRAS.522.3092L}).

In our model, stars with initial masses between 8--40 \mdot{} and $107.2$--\mbox{$203.4$ \mdot{}} are assumed to explode as core-collapse and pair-instability SNe, respectively. The latter corresponds to stars with final helium-core masses of $65$--\mbox{$133$ \mdot{}} that are expected to undergo thermonuclear runaway caused by the pair-production instability \citep{2002ApJ...567..532H}. In practice, the low-SFR galaxies considered in this study do not produce any pair-instability SN progenitors due to the locally mass-conserving IMF-sampling method. Core-collapse SNe are supplemented with thermal explosion energies of $10^{51}$ erg, and metal yields and remnant masses from \citet{2004ApJ...608..405C}, spatially distributed in a similar way as the stellar wind using \healpix{}. Stars between the range of 40 \mdot{} and \mbox{$107.2$ \mdot{}} are assumed to collapse to black holes without SNe. 

\subsubsection{Regularised integrator \ketju}

In order to accurately model the gravitational interactions at small separations between stars in the young star clusters without gravitational softening, we utilise the \ketju{} integration technique \citep{2017ApJ...840...53R} implemented originally in the \gadget{} code. The most recent version of \ketju{} \citep{2023MNRAS.524.4062M} and its regularised integrator library \mstar{} \citep{2020MNRAS.492.4131R} are publicly available\footnote{\href{https://www.mv.helsinki.fi/home/phjohans/ketju}{https://www.mv.helsinki.fi/home/phjohans/ketju}} as a module for the \textsc{gadget-4} code \citep{2021MNRAS.506.2871S}. The \gadget-\ketju{} code has been used in a wide range of studies examining the dynamics of supermassive black holes in isolated mergers of early-type galaxies \citep{2018ApJ...864..113R, 2019ApJ...872L..17R}, dark matter halos and low-mass galaxies \citep{2024MNRAS.532.4681P} and gas-rich galaxies \citep{2023MNRAS.520.4463L, 2024MNRAS.528.5080L, 2024MNRAS.530.4058L} as well as in cosmological simulations of forming early-type galaxies \citep{2021ApJ...912L..20M, 2022ApJ...929..167M}. We use the \gadget-version of \ketju{} as described in \citet{2021ApJ...912L..20M} as it is directly compatible with \sphgal. 

The key idea of \ketju{} is to select regions (hereafter \textit{\ketju-regions}) with radius $r_\mathrm{\ketju}$ up to few tens of parsecs in size within the \gadget{} simulation and to integrate the gravitational dynamics within the regions at high accuracy without the need for gravitational softening. The regularised integrator \mstar{} used in the regions is based on three key numerical methods that enable efficient and accurate integration as follows.
\begin{itemize}
    \item Algorithmic regularisation \citep{1999AJ....118.2532P, 1999MNRAS.310..745M, 2006MNRAS.372..219M, 2008AJ....135.2398M} using time-transformed equations of motion together with a custom leapfrog integrator for efficient non-softened integration of N-body systems avoiding the Newtonian coordinate singularity at small particle separations.
    \item Minimum spanning tree coordinate system \citep{2020MNRAS.492.4131R} to minimise floating-point round-off error and thus improve the integration accuracy.
    \item Gragg–Bulirsch–Stoer \citep{1965SJNA....2..384G, Bulirsch1966} extrapolation method to reach extremely high user-desired integration accuracy.
\end{itemize}
\mstar{} is efficiently parallelised and can treat up to thousands of simulation particles per regularised region. The library version of \mstar{} also includes an option for post-Newtonian equations of motion and a method for sophisticated integration order control \citep{2021ApJ...912L..20M}.

In previous studies the \ketju-regions have been placed around intermediate-mass or supermassive black holes in simulations with a stellar particle mass resolution of $\sim 0.08$ -- $10^5\,\mathrm{M}_\odot$. However, \ketju{} allows for placing the regions at arbitrary locations and centering on very massive objects is not required for the integration technique to work. In this study for the first time we use \ketju{} regions around massive stars in a star-by-star mass resolution dwarf galaxy simulation. More specifically, we include \ketju-regions around every star above a certain mass threshold $m_i$ to capture the collisional dynamics of these stars without gravitational softening. Outside the regions the code time integration is second-order accurate and all gravitational interactions are softened with the softening lengths $\epsilon$ of the particles.

As the \ketju-region size has to be at least $2.8$ times the gravitational softening length of the particles, we adopt here a factor of 3, i.e. $r_\mathrm{\ketju}=3\epsilon_*$ where $\epsilon_*$ is the stellar softening length. \ketju-regions that overlap are merged into one, thus the central regions of star cluster can host \ketju-regions that are significantly larger than the single region size. The dynamics of low-mass stars below the threshold mass $m_i$ remains softened unless the low-mass stars are within a regularised region of a nearby massive star. The reservoir stars (before IMF-sampling is done), old disk particles, gas particles and dark matter particles are also not included in the \ketju{} calculations. All interactions that are not between stars in a \ketju-region are computed using the standard leapfrog integrator and tree force algorithm in \gadget. We do not use the optional star-star softening inside the \ketju-regions since our N-body particles represent actual physical stars instead of macro-particles as has been the case in most previous \ketju{} studies. For the user-given accuracy parameters we use a Gragg-Bulirsch-Stoer tolerance of $\eta_\mathrm{GBS}=10^{-7}$ and a end-time iteration tolerance of $\eta_\mathrm{t}=10^{-3}$. The simulations in this study are fully Newtonian.

\subsubsection{Code comparison: \bifrost}

In this study we use the direct N-body code \bifrost{} for comparison star cluster simulations to verify the improved accuracy of our \gadget-\ketju{} star cluster dynamics. The benchmark code \bifrost{} \citep{2023MNRAS.522.5180R,2023MNRAS.521.2930R,Rantala2024a, 2024MNRAS.531.3770R} 
is a modern GPU-accelerated direct-summation N-body simulation code based on the hierarchical \citep{Rantala2021} fourth-order forward symplectic integrator \citep{Chin1997,Chin2005,Chin2007,Dehnen2017a}. Besides the forward integrator, \bifrost{} uses a number of secular and regularised few-body integration techniques closely related to \mstar{} \citep{2020MNRAS.492.4131R} to solve the gravitational dynamics of binary stars, close fly-bys, multiple systems and small clusters around massive black holes. In \bifrost{} no gravitational softening is used. For additional details of the code see \cite{2023MNRAS.522.5180R} and Appendix A of \cite{2024MNRAS.531.3770R}.

\subsection{Simulation setups and initial conditions}\label{section:ICs}
\subsubsection{N-body simulations}

In order to compare our \sphgal+\ketju{} code with the direct N-body code \bifrost, we run two sets of isolated star cluster simulations without stellar evolution. The cluster initial conditions have Plummer density profiles and the half-mass radii $r_{50\%}$ are set according to the observed mass-size relation of \citet{Marks2012}. Individual stellar masses are sampled from a Kroupa IMF, adopting the cluster mass-dependent upper mass limit to stellar masses according to \citet{Yan2023}, see also \citet{2024MNRAS.531.3770R} for details. The first set follows the evolution of a star cluster consisting of $\sim10^3$ stars and a cluster mass of \mbox{$\sim 600$ \mdot{}} with \mbox{$r_{50\%}=0.18$ pc}. The second set of simulations considers the evolution of a cluster with $\sim10^4$ stars and a mass of \mbox{$\sim6000$ \mdot} with \mbox{$r_{50\%}=0.28$ pc}. We created 10 random realisations of both initial conditions. The sample of runs is detailed in Tab. \ref{table:ics} and further information about the simulation parameters is given in Section \ref{section:parameters}. We label the simulations executed with \bifrost{} as \texttt{BIFROST}, simulations run while \ketju{} is enabled as \texttt{KETJU} and the simulations where \ketju{} is disabled as \texttt{GADGET}\footnote{The latter is named as such because there are no SPH-particles or stellar evolution in the pure N-body runs, and the computation of gravitational dynamics is done in \sphgal{} with the standard leap-frog integrator and tree force algorithm of \gadget.}.

\subsubsection{Dwarf galaxy simulations}

The low-metallicity ($Z=0.01$ \zdot) dwarf galaxy initial conditions are adopted from \citet{2023MNRAS.522.3092L}, based on the dwarf galaxy models of \citet{2016MNRAS.458.3528H}. We use the compact initial condition with virial mass of \mbox{$4\times 10^{10}$ \mdot}, including a \mbox{$4\times 10^7$ \mdot{}} gas disk and a \mbox{$2\times 10^7$ \mdot{}} old stellar disk with disk scale lengths of \mbox{0.73 kpc}. The initial mass-resolution is \mbox{4 \mdot{}} for gas and disk stars and \mbox{$\sim6800$ \mdot{}} for dark matter. The gravitational force-softening length is set to \mbox{62 pc} for dark matter, \mbox{0.1 pc} for gas and \mbox{0.1 pc} for pre-existing disk star particles. The smallest SPH smoothing lengths containing the 100 nearest gas particles are $\sim 0.3$ pc, corresponding to densities of \mbox{$10^5$ cm$^{-3}$} at the typical star formation threshold. For newly formed stars we use $\epsilon_*$ of \mbox{0.1 pc} or \mbox{0.01 pc} (see Section \ref{section:parameters} for details) except when for stars that are within \ketju-regions in the \ketju-simulations. The hydrodynamical simulations are named as \texttt{hyd\_KETJU} when \ketju{} is enabled and as \texttt{hyd\_SPHGAL} when \ketju{} is disabled.

\begin{figure*}
\includegraphics[width=\textwidth]{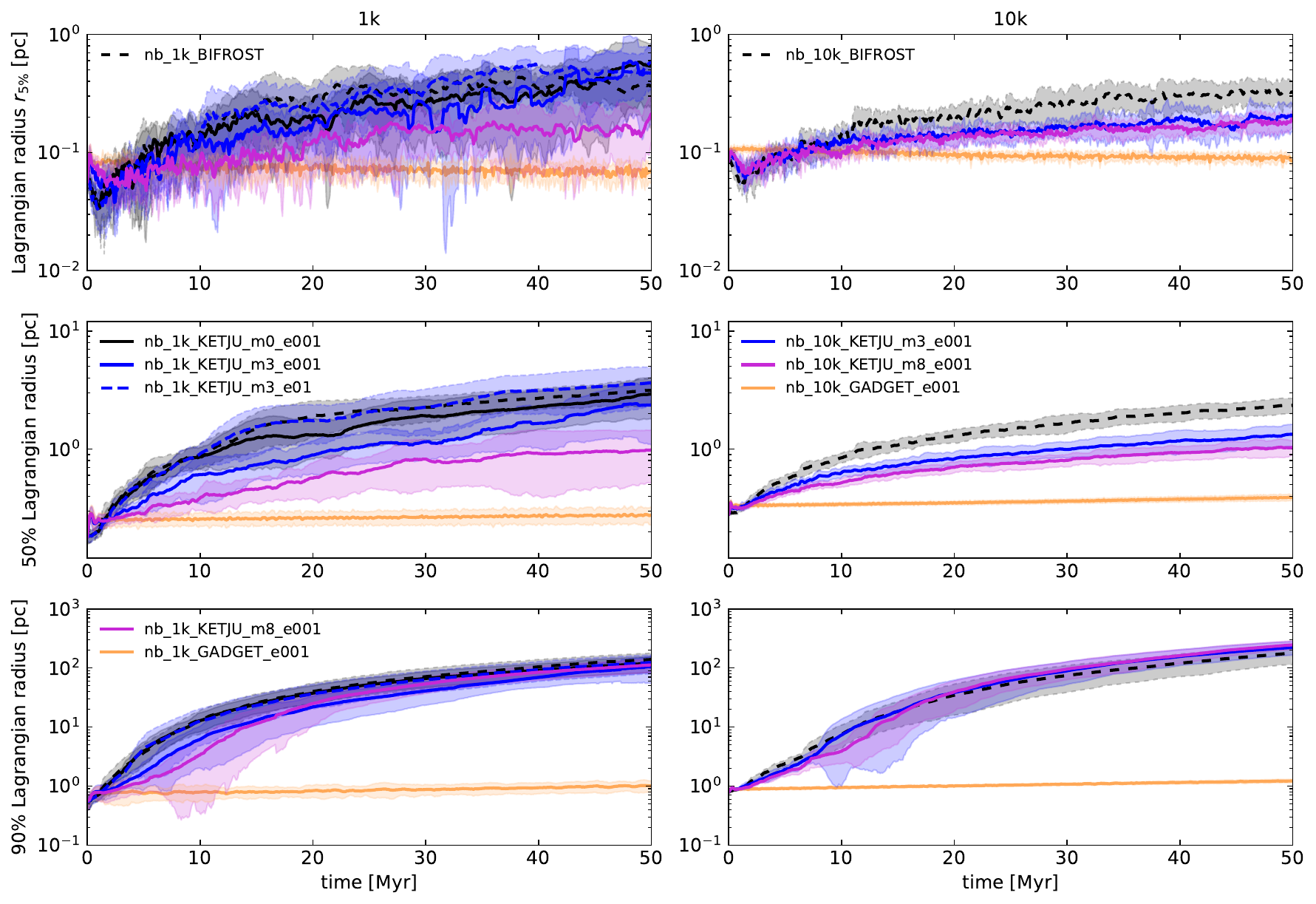}
\caption{The 5\% (\textit{top}), 50\% (\textit{middle}) and 90\% (\textit{bottom}) Lagrangian radii in the \texttt{1k} (\textit{left}) and the \texttt{10k} (\textit{right}) simulations, run in isolation without stellar evolution. The lines show the mean values of the sample of 10 runs per parameter combination, and the shaded regions show the standard deviations of the samples. The simulations on the left are coloured as follows: \texttt{nb\_1k\_BIFROST} in dashed black, \texttt{nb\_1k\_KETJU\_m0\_e001} in solid black, \texttt{nb\_1k\_KETJU\_m3\_e001} in blue, \texttt{nb\_1k\_KETJU\_m3\_e01} in dashed blue, \texttt{nb\_1k\_KETJU\_m8\_e001} in magenta and \texttt{nb\_1k\_GADGET\_e001} in orange, and equivalently for the \texttt{10k}-runs on the right. The sizes evolve increasingly similar to the direct N-body simulations (\bifrost) in simulations that solve increasing fractions of close gravitational interactions without softening. However, even a small fraction of accurately solved interactions (e.g. \texttt{nb\_1k\_KETJU\_m8\_e001} and \texttt{nb\_10k\_KETJU\_m8\_e001}) already drastically improves the cluster evolution compared to fully softened simulations (\texttt{nb\_1k\_GADGET\_e001} and \texttt{nb\_10k\_GADGET\_e001}). \label{fig:nbody_lagrangian}}
\end{figure*}

\subsection{Softening and regularisation parameters}\label{section:parameters}

We can estimate the desired minimum size for the \ketju{} region by calculating the typical strong encounter impact parameter or $90^\circ$ deflection radius $b_{90}$ \citep{2008gady.book.....B} for two equal mass stars $m_*$ as
\begin{equation}\label{eq:b90}
    b_{90} = \frac{2Gm_*}{V^2}
\end{equation}
with the encounter velocity $V\sim\sigma$, the stellar velocity dispersion. Another possible limit for the region size is the transition from a hard to a soft binary, defined as the limit where the potential energy of a binary equals the mean kinetic energy of a particle in the surrounding cluster, which for an equal-mass binary becomes
\begin{equation}\label{eq:abin}
    a_\mathrm{bin} \sim \frac{Gm_*^2}{\left< m_*\right>\sigma^2}, 
\end{equation}
where $a_\mathrm{bin}$ is the binary semi-major axis.  
As we are interested mainly in the evolution of massive stars, we can assume that they segregate rapidly to the central region of the cluster resulting in $\left< m_*\right>\sim m_*$ at the centre and yielding $a_\mathrm{bin} \sim b_{90}$.

The 1D central velocity dispersions of the $\sim 600$ \mdot{} and \mbox{$\sim 6000$ \mdot{}} clusters with Plummer profiles are $\sigma_\mathrm{0,1k}=1$ km s$^{-1}$ and $\sigma_\mathrm{0,10k}=2.6$ km s$^{-1}$, respectively. The largest stellar masses are $\sim15$ \mdot{} and $\sim55$ \mdot, respectively. Using Eqs. \ref{eq:b90} and \ref{eq:abin}, we get $b_{90,1k}\sim 0.06$ pc and $b_{90,10k}\sim 0.04$ pc. Compact ($r_{50\%}<1$ pc) bound young star clusters that form in the isolated dwarf galaxies have typical masses in the range from a few tens to $\sim900$ \mdot{} and typical 1D velocity dispersions of 0.1--2.5 km s$^{-1}$, resulting in typical $b_{90}$ values of less than $\sim0.1$ pc. 

Optimal $r_\mathrm{\ketju}$ values for strong gravitational interactions to be treated with the regularised \ketju{} integrator should thus be larger than 0.01--0.1 pc for our simulations. In addition to defining  which interactions are computed with the \ketju-integrator, $r_\mathrm{\ketju}$ also controls the computational cost of the integrator due to the scaling of the direct N-body problem with the particle number ($N$) as $\sim N^2$. We do not therefore wish to make the \ketju-regions larger than necessary. In an idealised environment of constant stellar density, the total $N$ in a \ketju-region scales as $N\propto r_\mathrm{\ketju}^3$. In case dense systems form, a large value of $r_\mathrm{\ketju}$ can lead to an increased computational load in \ketju\footnote{For a detailed description of the scaling properties of regularized integrators, see Section 5.2 of \citet{2017ApJ...840...53R}. But please note that the discussion considers an older version of \ketju, before updates to its perturber treatment \citep{2022ApJ...929..167M} and parallelisation \citep{2020MNRAS.492.4131R}.}. A large value of $r_\mathrm{\ketju}$ would additionally allow close \ketju-regions to be combined as one, further increasing the number of particles integrated in one region. We thus test values of $\epsilon_*$ of 0.01 pc and 0.1 pc in both N-body and hydrodynamical runs, which define $r_\mathrm{\ketju}$ as 0.03 pc or 0.3 pc, respectively. These \ketju-regions sizes bracket the optimal values estimated using Eqs. \ref{eq:b90} and \ref{eq:abin}.

We run a sample of ten random initialisations of both idealised cluster initial conditions ($10^3$ and $10^4$ stars) with \bifrost; with \sphgal{} (that is, without hydrodynamics or stellar evolution, the standard \gadget) using a $\epsilon_*=0.01$ pc; and with \sphgal+\ketju{} using $\epsilon_*=0.01$ pc and initiating \ketju{} integration within $r_\mathrm{\ketju}=$ \mbox{$0.03$ pc} either around all $m_i>3$ \mdot{} or around all $m_i>8$ \mdot{} stars. In addition, we run the \texttt{1k} initial conditions with \sphgal+\ketju{} using $r_\mathrm{\ketju}=0.3$ pc around $m_i>3$ \mdot{} stars and $r_\mathrm{\ketju}=0.03$ pc around \textit{all} stars. We do not execute the \texttt{10k}-sample with \mbox{$r_\mathrm{\ketju}=0.3$ pc} because it includes initially more than $5700$ stars in the central \mbox{0.3 pc}, and running it with the larger \ketju-region size is prohibitively expensive for tens of Myrs \citep{2020MNRAS.492.4131R}. The simulations are run for 50 Myr, which is significantly longer than the half-mass relaxation times of \mbox{$1.4$ Myr} and \mbox{$5.7$ Myr} of the \texttt{1k} and \texttt{10k} initial conditions, respectively. The labels and parameter details of the simulations are collected in Tab. \ref{table:ics}.

In the hydrodynamical runs, we test both $\epsilon_*=0.1\, \mathrm{pc}$ and $\epsilon_*=0.01\, \mathrm{pc}$, i.e. $r_\mathrm{\ketju}=0.3$ and $r_\mathrm{\ketju}=0.03$ pc, and run the simulations with and without \ketju{}. As clusters more massive than \mbox{$\sim100$ \mdot{}} are practically guaranteed in our simulations to form with stars that are more massive than 3 \mdot{} \citep[see e.g.][Fig. 14]{2023MNRAS.522.3092L}, we initiate the regions around $m_i>3$ \mdot{} stars to allow the majority of the massive clusters to have at least one \ketju-region for part of their early evolution. \ketju{} is triggered in total for $\sim7000$ stars during both regularised hydrodynamical simulations and the maximum number of stars within the individual regularised regions is 1000--2000 in \texttt{hyd\_KETJU\_m3\_e01} and 100--200 in \texttt{hyd\_KETJU\_m3\_e001}.

\subsection{Identification of star clusters}
In the hydrodynamical dwarf galaxy simulations, we identify bound star clusters using \textsc{friends-of-friends} and \textsc{subfind} \citep{2001MNRAS.328..726S, 2009MNRAS.399..497D}, which are structure finding algorithms included in \gadget. In order to follow the long-term evolution of clusters especially in the \ketju-runs, we use a linking length of 1 pc to recover the extended distribution of stars. The \textsc{subfind}-algorithm has been modified to account for the non-softened gravitational potential within the \ketju{}-regions in the runs where \ketju{} is enabled. We require at least 50 bound stars to be counted as a cluster, and exclude clusters with a mean stellar age younger than 1 Myr when making direct comparisons to observations where deeply embedded, extremely young clusters are often excluded. The cluster finding is performed for each snapshot over a timespan of 500 Myr in 1 Myr intervals.

\section{Comparison of KETJU with direct N-body}\label{section:nbody_results}

The number of interactions integrated with \ketju{} depends on the number and sizes of the regularised regions. Stars with initial masses of $m_i>3$ \mdot{} and $m_i>8$ \mdot{} account typically for $\sim 2\%$ and $0.5\%$ of all stars in the idealised cluster, respectively. With \mbox{$r_\mathrm{\ketju}=0.3$ pc} and \mbox{$m_i=3$ \mdot{}} in \texttt{nb\_1k\_KETJU\_m3\_e01}, initially all interactions between stars inside $r_{50\%}$ are integrated with \ketju{} as one combined \ketju-region can cover the entire volume within $r_{50\%}$. The fraction of regularised gravity interactions reduces as mass-segregation proceeds and the clusters expand, dropping to a few tens of per cent during the first 10--20 Myr of evolution. When the size of the regularised region is reduced to $r_\mathrm{\ketju}=0.03$ pc, the fraction of regularised interactions within $r_{50\%}$ is initially $\sim 10\%$ and $\sim 20\%$ in the \texttt{nb\_1k\_KETJU\_m3\_e001} and \texttt{nb\_10k\_KETJU\_m3\_e001} runs, respectively. The fraction reduces gradually to a few per cent as massive stars are segregated and/or ejected out of the cluster (since stellar evolution is switched off here). The fraction of stars within \ketju-regions is close to the fraction of $m_i>3$ \mdot{} stars, indicating that most of these stars have moved onto more distant orbits with low stellar densities or escaped the cluster and evolve in isolation. When $m_i$ is increased to 8 \mdot, the \texttt{nb\_1k\_KETJU\_m8\_e001}-run has only a handful of stars above the mass threshold. The initial fraction of stars within \ketju-regions is (averaged over 10 runs) only a few per cent and plateaus at half a per cent during the simulations, close to the fractions of $m_i>8\, \rm M_{\odot}$ stars. The \texttt{nb\_10k\_KETJU\_m8\_e001}-run has a $\sim 7\%$ initial fraction of regularised interactions within $r_{50\%}$, which drops to 1--2\% after 10 Myr of evolution. 

As the massive stars segregate to the cluster centre, they can form binaries, undergo strong gravitational two or few-body interactions and gradually get ejected from the clusters. Low-mass clusters simulated with \ketju{} can thus end up being integrated with fully softened gravity if all of the relatively rare massive stars escape. In this case, the size evolution of the cluster halts entirely. This is essentially the cause of the plateau in the inner Lagrangian radii after 25 Myr in the \texttt{nb\_1k\_KETJU\_m8\_e001}-sample discussed in Section \ref{section:lagrangian}.

\subsection{Lagrangian radii}\label{section:lagrangian}

Fig. \ref{fig:nbody_lagrangian} shows the evolution of the Lagrangian radii enclosing 5\%, 50\% and 90\% of the total mass of the clusters for the N-body simulations during 50 Myr. The \texttt{nb\_1k\_BIFROST} and \texttt{nb\_10k\_BIFROST} runs are the benchmarks against which the \sphgal{} and \sphgal{}+\ketju{} simulations are compared to. 

The fully softened simulations \texttt{nb\_1k\_GADGET\_e001} and \texttt{nb\_10k\_GADGET\_e001} show practically no evolution in the Lagrangian radii, especially when compared to the (partially) collisional simulations. Clusters in the softened gravity would only lose mass if the clusters were exposed to an external tidal field. To check whether changing the softening length impacts the cluster size, we ran the \sphgal-simulations without \ketju{} using a larger $\epsilon_*=0.1$ pc softening and saw no changes compared to the $\epsilon_*=0.01$ pc run. We thus omit the simulation with $\epsilon_*=0.1$ from further analysis.

The clusters in the simulations with regularisation start compact and their centres contract even further during the first few Myr while mass-segregation proceeds. In response, the outer regions expand, up to a factor of 10 and 100 for the 50\% and 90\% radii, respectively. The inner regions of individual  \texttt{nb\_10k\_BIFROST}-runs and \texttt{nb\_10k\_KETJU}-runs go through cycles of gradual contraction (see Fig. \ref{fig:appendix1} in Appendix \ref{section:appendix1}) that is halted by binary formation and rapid expansion of $r_{5\%}$. The expansion is then finished through the ejection of massive star(s) from the centre. However, this is not visible after the first core collapse cycle (around $\sim1$ Myr in the \bifrost{} run) when the cluster size is averaged over 10 runs. This is because each random simulation realisation has a different time-scale for the collapse-cycle depending on the masses of the most massive stars still present in the cluster. Overall, the mean 5\% Lagrangian radius increases after the first few Myr, as the clusters have too low masses to undergo further core collapse (see Fig. \ref{fig:appendix1} and e.g.  \citealt{1989MNRAS.237..757H}). 

The simulations that either integrate all close interactions with \ketju{} (\texttt{nb\_1k\_KETJU\_m0\_e001}) or include a larger fraction of the cluster stars in the \ketju{} integration (\texttt{nb\_1k\_KETJU\_m3\_e01}) evolve almost identically to the direct N-body runs. The results of the other simulations that use \ketju{} but only include stars above a mass threshold as the \ketju{} region centres (\texttt{KETJU\_m3}, \texttt{KETJU\_m8}) are inbetween the direct N-body results and the fully softened runs. This signifies that the dynamical processes important for the internal evolution of the clusters, such as mass-segregation and core collapse, are captured significantly better in all runs with \ketju{} compared to fully softened runs. Comparing \texttt{nb\_10k\_KETJU\_m3\_e001} and \texttt{nb\_10k\_KETJU\_m8\_e001}, we see that the dynamical evolution of the clusters is not very sensitive to the exact number of \ketju-regions. In other words, even only accounting for the collisional evolution of a few per cent of the stars in the cluster results in significantly more realistic cluster evolution.

\begin{figure*}
\includegraphics[width=0.9\textwidth]{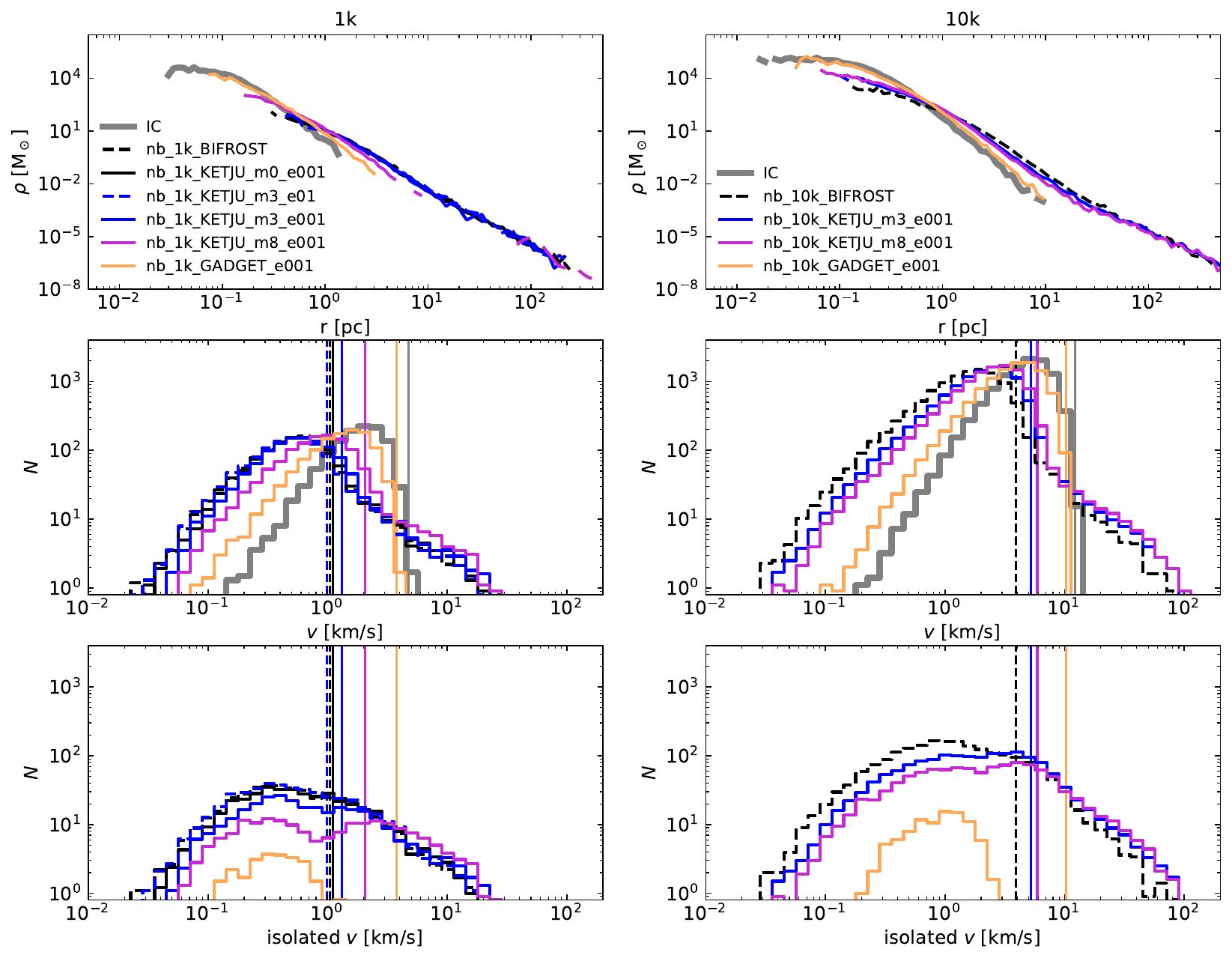}
\caption{The radial stellar density (\textit{top}), the stellar velocity distribution (\textit{middle}) and the velocity of isolated stars (zero stars within a radius of 1 pc; \textit{bottom}) in the final snapshot (50 Myr) of the \texttt{1k} (\textit{left}) and \texttt{10k} (\textit{right}) N-body runs. The line colours are as in Fig. \ref{fig:nbody_lagrangian} and the lines show the mean values of the sample of 10 runs per parameter combination. The density and velocity distribution in the initial conditions is shown in gray thick lines in the top and middle rows, and the initial escape velocity within $r_{50\%}$ is plotted with a thin gray vertical line in the middle panels. The escape velocity within $r_{50\%}$ in the final snapshot of each run is indicated with the vertical line in the same linestyle. Star clusters evolved with \bifrost{} or \ketju{} evolve toward energy equipartition, expand, and produce escaping stars, while clusters simulated with fully softened gravity evolve very little from the initial condition.\label{fig:nbody_density}}
\end{figure*}

\subsection{Stellar density and velocity}

Fig. \ref{fig:nbody_density} shows the density and velocity distributions of stars in the N-body simulations at 50 Myr, together with the initial conditions. We indicate the escape velocity at $r_{50\%}$,
\begin{equation}
    v_\mathrm{esc}=\sqrt{2|\Phi_{50\%}|},
\end{equation}
where $\Phi_{50\%}$ is the gravitational potential at $r_{50\%}$, in the initial conditions and at 50 Myr. The sample-averaged $v_\mathrm{esc}$ is between \mbox{$\sim1$--$5$ km s$^{-1}$} for the \texttt{1k}-runs and $\sim4$--$13$ \mbox{km s$^{-1}$} for the \texttt{10k}-runs. We also show separately stars that are isolated by at least \mbox{1 pc} from any other star, which comprise walkaway (\mbox{$10$ km s$^{-1} <$ } $v$ \mbox{$<30$ km s$^{-1}$}) and runaway ($v$ \mbox{$>30$ km s$^{-1}$}) stars with velocities larger than the typical escape velocity of star clusters. Some of the stars have large velocities as they reside in binaries. Thus, any binary members are excluded from the isolated star analysis to avoid confusing binary motion with escaping stars. As stellar evolution is not included in the N-body comparison runs, escapes e.g. due to exploding binary companion stars do not occur, and the stars reach their escape velocities purely due to few-body interactions.

As with the Lagrangian radii, the \texttt{nb\_1k\_GADGET\_e001} and the \texttt{nb\_10k\_GADGET\_e001} runs do not evolve much from the initial condition, neither in the density nor the velocity distribution. The density distribution expands very little, and the stars maintain almost the same velocity distributions throughout the 50 Myr of evolution. As two-body interactions are softened in the dense central region of the cluster, energy equipartition cannot proceed, mass-segregation is suppressed and no escaping stars are produced due to the lack of strong few-body interactions. 

\begin{figure*}
\includegraphics[width=\textwidth]{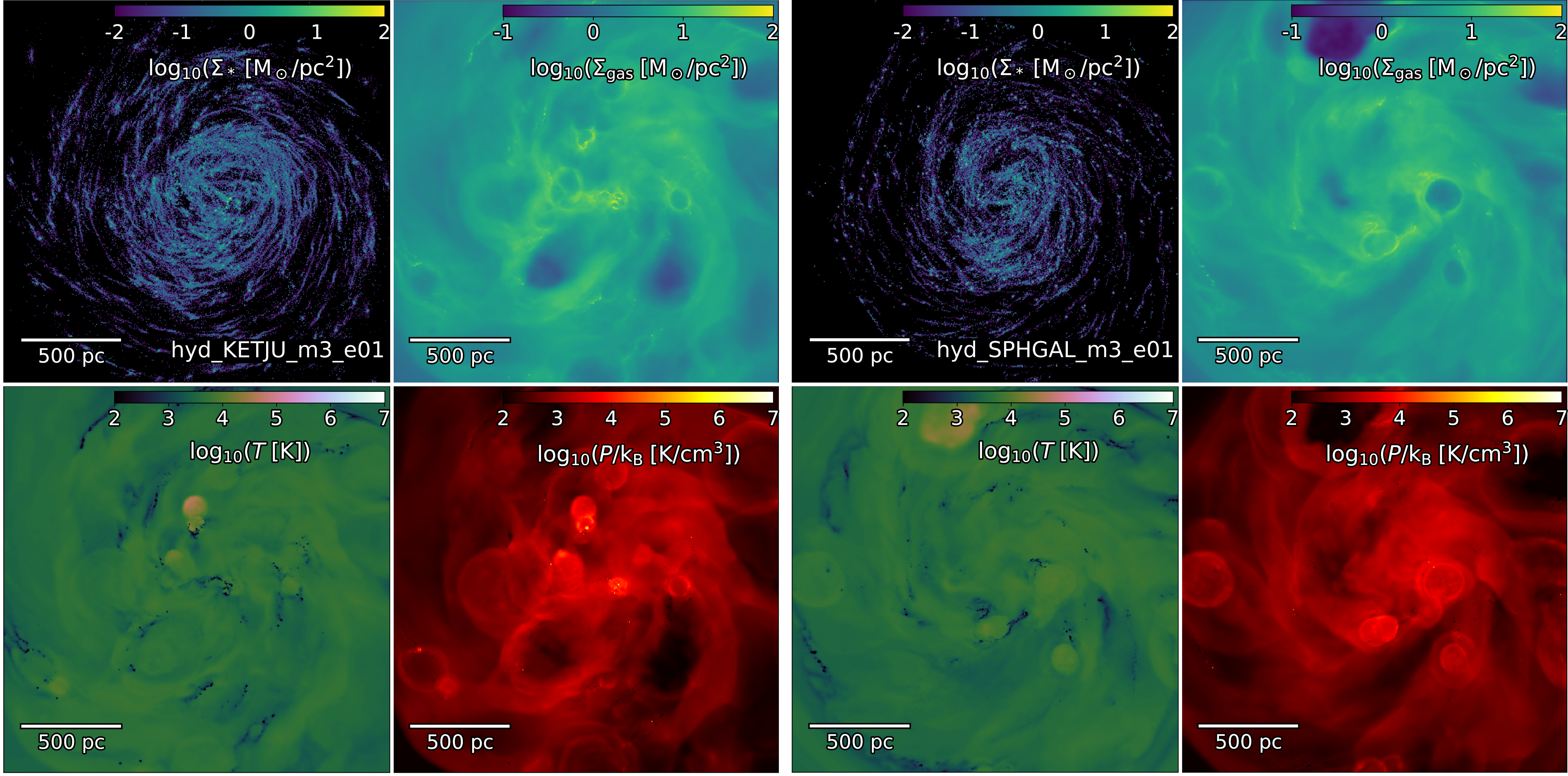}
\caption{The stellar and gaseous surface density (\textit{top left} and \textit{right}), the gas temperature (\textit{bottom left}) and the thermal gas pressure (\textit{bottom right}) in the \texttt{hyd\_KETJU\_m3\_e01} (four left panels) and \texttt{hyd\_SPHGAL\_m3\_e01} (four right panels) simulations at 500 Myr. The tidal tails of star clusters that are losing mass can be seen as leading and trailing arms around the concentrations of stellar mass on the left, while clusters on the right remain compact and only lose a little mass in tidal tails. SN-bubbles and photoionised regions are visible in the gaseous distribution in both simulations. The field of view is 2 kpc and the image resolution is $\sim 4$ pc per pixel.
\label{fig:2D}}
\end{figure*}

On the other hand, the simulations that do capture at least some of the close gravitational encounters show significant expansion during the evolution of the clusters (top panels of Fig. \ref{fig:nbody_density}). During \mbox{50 Myrs} of evolution, the central densities decrease and the outer regions expand. The effect is stronger in the runs that treat a larger fraction of gravitational interactions accurately without softening. As we saw in Fig. \ref{fig:nbody_lagrangian}, the expansion of the clusters is slower in the \ketju-runs compared to \bifrost, because interactions between low-mass stars outside the regularised regions are still softened and their segregation is thus somewhat suppressed. 

As mass-segregation proceeds, interactions lead to the redistribution of stars onto more distant (less bound) orbits with lower average orbital velocities while a handful of stars reach velocities above the escape velocity. Since the strongest interactions occur in the cluster centre and often include massive stars, we see more consistent numbers of escaping stars being produced in simulations that have at least some regularised interactions (\bifrost-runs and \ketju-runs). The runs with \ketju-regions around all stars (\texttt{nb\_1k\_KETJU\_m0\_e001}) or in a larger volume (\texttt{nb\_1k\_KETJU\_m3\_e01}) show almost identical results to the \texttt{nb\_1k\_BIFROST} run, both in density and in the velocity distributions. The runs in the \texttt{nb\_10k\_BIFROST}-sample, on the other hand, have $\sim 90$ stars with velocities larger than \mbox{10 km s$^{-1}$} at 50 Myr, compared to the equivalent value of $\sim 120$ and $\sim 130$ in the \texttt{nb\_10k\_KETJU\_m3} and \texttt{nb\_10k\_KETJU\_m8} samples. The cause for the larger number of runaway stars in the \ketju{} simulations are the higher central densities that enable an elevated number of strong gravitational interactions. Here it is worth stressing again that the clusters evolved entirely with softened gravity produce practically no escaping stars due to the suppression of strong gravitational encounters, even tough they have the highest stellar densities for the majority of the simulation time.

\begin{figure}
\includegraphics[width=\columnwidth]{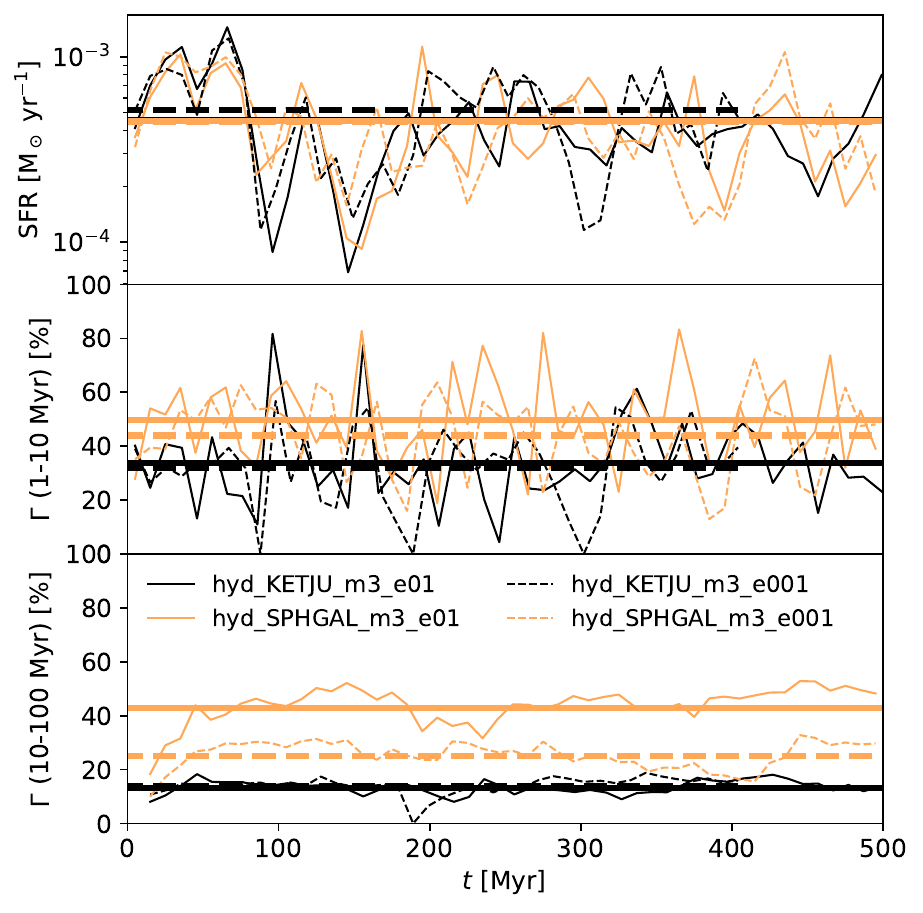}
\caption{The SFR averaged in 10 Myr bins (\textit{top}), and $\Gamma$ computed as the ratio between the CFR and SFR for bound star clusters between ages of 1--10 Myr (\textit{middle}) and 10--100 Myr (\textit{bottom}). The thin lines show the time variation for \texttt{hyd\_KETJU\_m3\_e01} (black solid), \texttt{hyd\_KETJU\_m3\_e001} (black dashed), \texttt{hyd\_SPHGAL\_m3\_e01} (orange solid) and \texttt{hyd\_SPHGAL\_m3\_e001} (orange dashed), and the thick horizontal lines show the respective time-averages.
\label{fig:sfr}}
\end{figure}

\section{Star formation and cluster evolution on galactic scales}\label{section:hyd_results}

Next we turn to the full hydrodynamical galaxy-simulations that follow the formation, evolution and disruption of star clusters in the evolving tidal field of a low-metallicity dwarf galaxy.
In Fig. \ref{fig:2D} we show the surface density maps of stars and gas, as well as the gas temperature and thermal pressure, at the end of the hydrodynamical \texttt{hyd\_KETJU\_m3\_e01} and \texttt{hyd\_SPHGAL\_m3\_e01} simulations  at \mbox{500 Myr}. The gaseous distributions are fairly similar in the two runs, given that they are sensitive to the temporal stochasticity of the feedback produced by young stars mainly on small spatial scales. Bubbles of photoionisation and SNe (e.g. the red blobs in the temperature panels) are visible in all gaseous quantities. The \sphgal-panels show good examples of SNe that have exploded within photoionised bubbles that are visible as concentric expansion fronts especially in the pressure panel.

On the other hand a clear difference between the two simulations can be seen in the distribution of stars. The regularised \ketju-simulation shows clear signs of cluster disruption through strong leading and trailing tidal tails associated with individual star clusters. The stellar streams with no clear concentrations of stellar mass are fully disrupted clusters. Some mass-loss is seen in the \sphgal-run as well, enabled by the cluster interacting with the tidal field. The clusters, however, still remain very compact when all gravitational forces are softened \citep[see][for a detailed discussion]{2022MNRAS.509.5938H}.

\subsection{Star formation rate and cluster formation rate}

We first inspect the impact of adding collisional stellar dynamics to hydrodynamical galaxy-simulations by comparing the global properties of the simulated galaxies in the \sphgal{} and \sphgal+\ketju{} runs. Fig. \ref{fig:sfr} shows the galaxy-wide SFRs, as well as the star cluster formation efficiencies $\Gamma$ as defined by the ratio between the cluster formation rate (CFR) and the SFR measured over a similar timespan. The SFRs shown in the top panel are averages over the past 10 Myr, accompanied by the corresponding overall average values. CFRs are computed using the total mass in young ($1$--\mbox{$10$ Myr}, middle panel) or intermediate age ($10$--\mbox{$100$ Myr}, bottom panel) bound clusters more massive than \mbox{$100$ \mdot{}}, and are then divided by the average SFR over 10 Myr and 100 Myr, respectively, to arrive at the value of $\Gamma$. We show both the \texttt{e01} and \texttt{e001} versions of the hydrodynamical simulations in Fig. \ref{fig:sfr}, and conclude that there are very little differences in the SFRs (top panel) between the runs with and without \ketju{} or in the runs with varying $\epsilon_*$, as indicated by the near identical time-averages over the 400--500 Myr of galaxy evolution. 

The value of $\Gamma$ (middle panel), on the other hand, decreases with the inclusion of more accurate stellar dynamics. As already indicated in Fig. \ref{fig:nbody_density}, collisional interactions cause a more rapid expansion and thus faster tidal disruption of clusters. Interactions with gas clouds and other clusters can also cause mass-loss events. Some of this evolution is captured in the \sphgal{} simulations as well, as shown already in \citet{2023MNRAS.522.3092L} where star clusters modelled with the standard \sphgal{} lost tens of per cents of their mass over hundreds of Myrs even with a \mbox{0.1 pc} gravitational softening.

\begin{figure*}
\includegraphics[width=\textwidth]{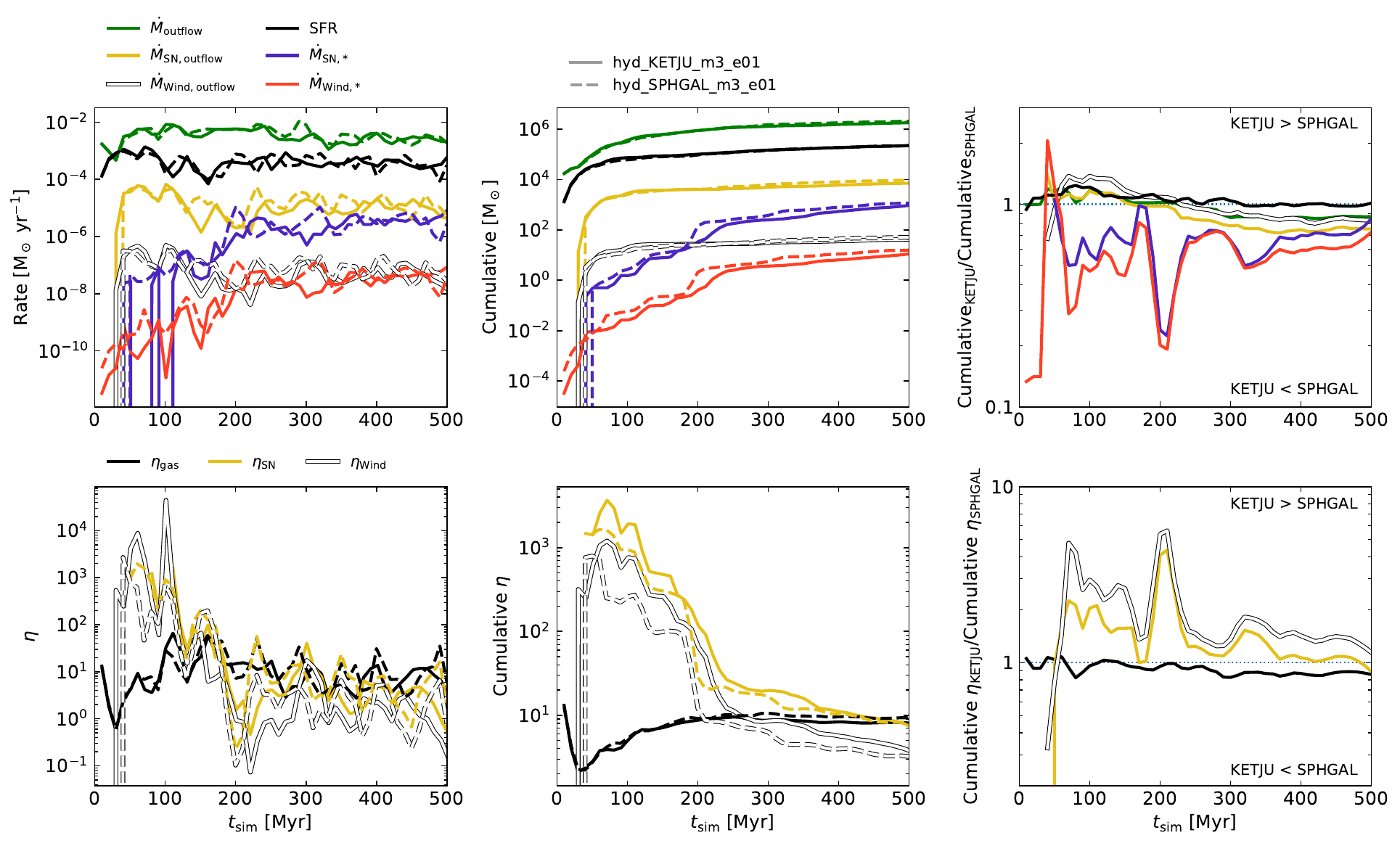}
\caption{\textit{Top}: the galactic outflow rate of gas, SN-material and wind-material, and the rate at which wind and SN-material are locked into new stars (\textit{left}) and the respective cumulative values (\textit{middle}). The lines from top to bottom are $\dot{M}_\mathrm{outflow}$, SFR, $\dot{M}_\mathrm{SN,outflow}$, $\dot{M}_\mathrm{SN,*}$, $\dot{M}_\mathrm{Wind,outflow}$ and $\dot{M}_\mathrm{Wind,*}$ (see text for details). The \texttt{hyd\_KETJU\_m3\_e01} and \texttt{hyd\_SPHGAL\_m3\_e01} simulations are shown in solid and dashed line style. The \textit{right} column shows the ratios between the cumulative masses in the \ketju-run and the \sphgal-run. The dashed horizontal line indicates the value of one. \textit{Bottom}: the mass-loading factor of gas (black lines), SN-material (yellow lines) and wind-material (white lines). The panels show the rates and loading factors (\textit{left}), the cumulative masses and loading factors (\textit{middle}) based on the cumulative values from the top middle panel, and the ratios between the cumulative loading factors in the \ketju{} and \sphgal{} simulations (\textit{right}). 
\label{fig:ofr}}
\end{figure*}

In Fig. \ref{fig:sfr}, the \sphgal{} runs have time-averaged $\Gamma$ (1--10 Myr) of 40--50\%, while the \ketju{}-runs result in $\Gamma$ (1--10 Myr) $\sim 30$--$35$\%. The global stellar mass and the number of gravitationally bound star clusters formed in the simulations are almost identical between the \sphgal{} and \ketju{} runs, but the \ketju-clusters begin losing bound stars dynamically immediately at formation. This is in addition to the stellar wind mass-loss, which happens in the \sphgal-run. The values of CFR decrease when averaged over 1--10 Myr, reducing also the \mbox{$10$ Myr} averaged $\Gamma$ when collisional dynamics are included. We will discuss the evolution of the individual clusters in more detail in Section \ref{section:clusters}. We note the persistent variation of $\Gamma$ from snapshot to snapshot caused by stochasticity in star formation that is active in only a few regions of the galaxy at any given time. Observations of $\Gamma$ in dwarf galaxies show similar variations from galaxy to galaxy, with sample mean or median values of 10--40\% and a standard deviation up to 20\% in young (1--10 Myr) clusters \citep{2023MNRAS.519.3749C, 2023ApJ...949..116C}. However, more important for the current study is the change of $\Gamma$ with cluster age; observed values of $\Gamma$ drop to a few per cent when cluster formation in the age range of 10--100 Myr is analysed. In Fig. \ref{fig:sfr}, only the \ketju{}-simulations recover 10\%-level values of $\Gamma$ for older clusters, caused by more efficient mass-loss.

The clusters in the idealised \texttt{nb\_1k\_GADGET\_e001} and \texttt{nb\_10k\_GADGET\_e001} runs showed no size-evolution in Fig. \ref{fig:nbody_lagrangian}. In the bottom panels of Fig. \ref{fig:sfr} we however see that the impact of the galactic tidal field in the hydrodynamical simulations on the dynamical mass-loss of the clusters is stronger when a smaller softening length is used in \texttt{hyd\_SPHGAL\_m3\_e001}. The \sphgal-run with a smaller $\epsilon_*$ shows some reduction in $\Gamma$ especially at older cluster ages caused by cluster expansion and mass-loss. The evolution is, however, still too slow to reach mass-loss rates equivalent to the \ketju-runs. Based on Fig. \ref{fig:sfr} we conclude that the evolution of clusters in the softened galaxy-scale \sphgal-simulations is sensitive to the specific adopted value of the softening length. The opposite is true for the \ketju-simulations where an order of magnitude change in the \ketju-region size leads to qualitatively similar results.

In the following, we will therefore concentrate on analysing the differences between the \texttt{hyd\_SPHGAL\_m3\_e01} and \texttt{hyd\_KETJU\_m3\_e01} runs.

\begin{figure}
\includegraphics[width=\columnwidth]{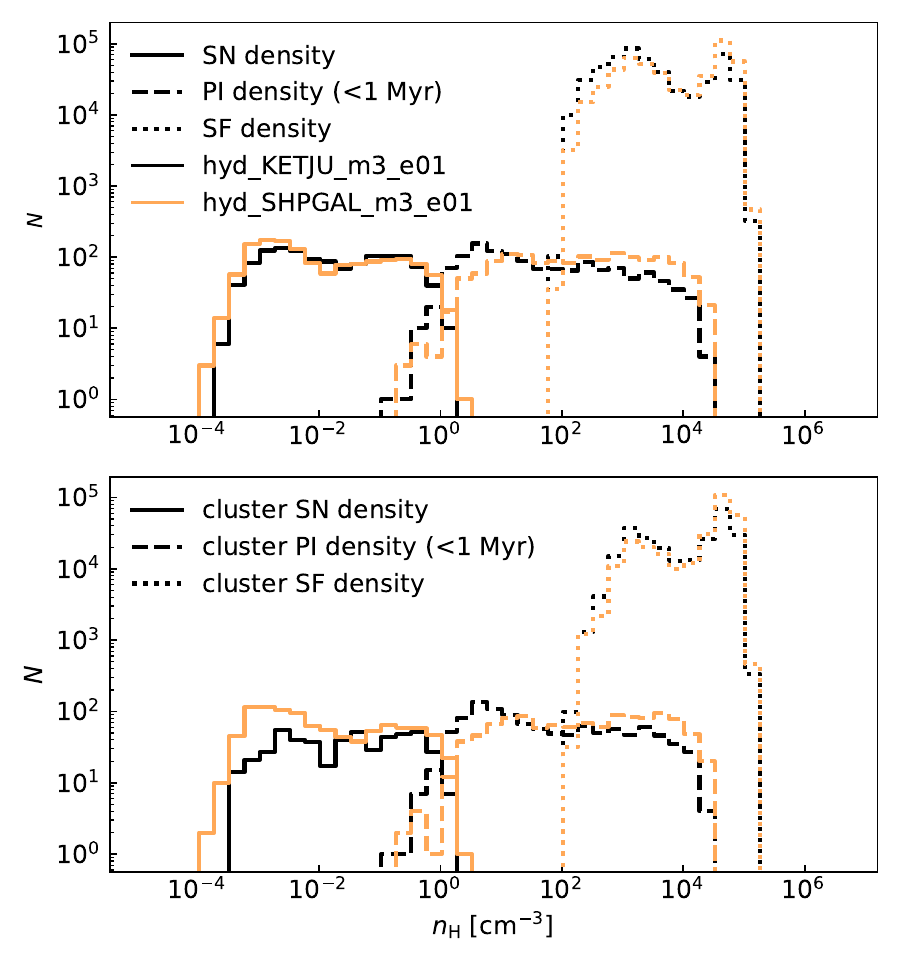}
\caption{ISM density around stars younger than 1 Myr (dashed) and at the location of SN events (solid), compared to the star formation density (dotted). The \textit{top} panel shows the values for all stars and the \textit{bottom} panel shows the values for stars in bound star clusters. The photoionisation and SN densities have been stacked in 1 Myr steps. The \texttt{hyd\_KETJU\_m3\_e01} and \texttt{hyd\_SPHGAL\_m3\_e01} simulations are shown in black and orange, respectively.
\label{fig:fb}}
\end{figure}

\subsection{Chemical enrichment and galactic outflows}

 The gas and stellar particles include information about how much each stellar feedback process contributed to the chemical composition of the particle. This enables us to trace the propagation of the wind and SN-material throughout the galaxy.
In Fig. \ref{fig:ofr} we show the time-evolution of the outflow properties and mass-loading factors in the \texttt{hyd\_KETJU\_m3\_e01} and \texttt{hyd\_SPHGAL\_m3\_e01} simulations. The outflows have been measured at a height of \mbox{1 kpc} above and below the galaxy midplane in \mbox{100 pc} thick slabs\footnote{See Section 5.4.1 in \citealt{2023MNRAS.522.3092L} for more details and further discussion regarding our definition of the mass-loading.}. We show the total gas outflow rate ($\dot{M}_\mathrm{outflow}$); outflow rate of SN and wind-material ($\dot{M}_\mathrm{SN,outflow}$ and $\dot{M}_\mathrm{Wind,outflow}$); the global SFR; and the locking rate of SN and wind-material into new stars ($\dot{M}_\mathrm{SN,*}$ and $\dot{M}_\mathrm{Wind,*}$). The loading factors $\eta$ have been computed as the ratio between the respective rate of outflow and star formation, which removes the material from the ISM and locks it into new stars: $\eta_\mathrm{gas}=\dot{M}_\mathrm{outflow}/\mathrm{SFR}$, $\eta_\mathrm{SN}=\dot{M}_\mathrm{SN,outflow}/\dot{M}_\mathrm{SN,*}$ and $\eta_\mathrm{Wind}=\dot{M}_\mathrm{Wind,outflow}/\dot{M}_\mathrm{Wind,*}$. We also compute the cumulative values to compare the total masses in  formed stars, the total chemical enrichment and the total mass-outflow out of the galaxy in the two simulations.

Based on Fig. \ref{fig:ofr}, the efficient cluster dissolution in such a low-SFR galaxy has only a minor impact on the galaxy-wide baryon cycle. The total mass in outflows, formed stars and locked metals show similar values in the \ketju{} and \sphgal{} runs to within a few tens of per cent. The massive stars in the \ketju{} simulation are less clustered and deposit their wind and SN-material in a larger ISM volume. This leads to periods of time when a lower fraction of wind and SN-material is being locked in new stars. The more efficient disruption of star clusters thus results in a locally lower recycling efficiency of massive star ejecta. This is also reflected in the mass-loading of these two components, which is overall larger in the \ketju-run. As time progresses, the enriched material injected outside of star clusters gets mixed in the galactic ambient ISM and other star-forming regions. The wind and SN-material locking rates and loading factors of the \ketju{}-simulation thus approach the \sphgal{}-run over a timescale of several 100 Myr. The sub-resolution method for metal-diffusion introduced in \citet{2013MNRAS.434.3142A} is not enabled here, thus the mixing of metals is the result of turbulent mixing of the particles (see the discussion in \citealt{2024arXiv240714599S} on the impact of metal-diffusion in Lagrangian codes).

Our \sphgal+\ketju{} simulation method can be used to model the galactic-scale impact of runaway massive stars that are ejected from star clusters due to their dynamical evolution. Previous high-resolution galaxy simulations have also explored the impact of runaway stars on the star formation and outflow properties of galaxies \citep{2020MNRAS.494.3328A, 2023MNRAS.526.1408S, 2023MNRAS.521.2196A}, however using ad-hoc sub-resolution models to give the stars their "excess" velocities. Notably, the models give a velocity kick to  massive stars immediately when they form. This essentially means that the stars are almost instantly removed from the star-forming regions and instead injected into the less dense ISM when their pre-SN feedback channels have just become active. With a typical escape velocity of 10 km $s^{-1}$, stars travel at least 10 pc from the cluster during the first Myr\footnote{The minimum velocity kick imposed in the sub-resolution models is often 3 km s$^{-1}$.}, therefore their critical role in regulating the gas-content of the clusters (see Section \ref{section:gas_expulsion}) may be fundamentally impacted by such "primordial" kicks. In collisional N-body models, the escape of the stars instead occurs gradually as the stars segregate, form binaries and receive kicks in dynamical interactions. Compared to the previous studies that found increased outflow rates and both increased and decreased star formation histories associated with runaway stars, our simulations show almost identical mass-outflow rates and star-formation histories with and without efficient star cluster disruption. The only notable change here is seen in the recycling of the metal-enriched material in newly formed stars, which is reduced due to the less clustered feedback. Our galaxies, however, also have lower metallicites and lower global SFRs compared to the previous studies, therefore a direct comparison is not possible.

\subsection{SN clustering}

In Fig. \ref{fig:fb} we show the ISM densities around young stars at formation, around photoionising young stars ($<1$ Myr) and around stars that have exploded as SN. To inspect how the accurate stellar dynamics impacts clustering of stellar feedback, we also show separately feedback events that occur in bound star clusters (bottom panel) recovered in both simulations (\texttt{hyd\_KETJU\_m3\_e01} and \texttt{hyd\_SPHGAL\_m3\_e01}). While photoionisation and star formation occur at slightly lower densities in the \ketju-run, the SN environmental densities are quite similar in both runs. This manifestly shows the impact of effective early stellar feedback in regulating the gaseous densities in star-forming regions. The only significant difference in the two runs is the number of SNe in clusters, which is consistent with Fig. \ref{fig:ofr} where we saw a lower recycling efficiency of wind and SN-material. The number of SNe in bound clusters reduces from 993 to 509, out of 1428 and 1273 SNe in total, respectively. The fraction of SNe in clusters therefore drops from $70\%$ in the \sphgal-simulation to $40\%$ in the \ketju-run. The effect of lowered SN-clustering on the galaxy outflows was however very minor as shown in Fig. \ref{fig:ofr}, with a difference of only 10\% in the total gas mass that escapes the galactic disk between the two simulations.

\begin{figure}
\includegraphics[width=\columnwidth]{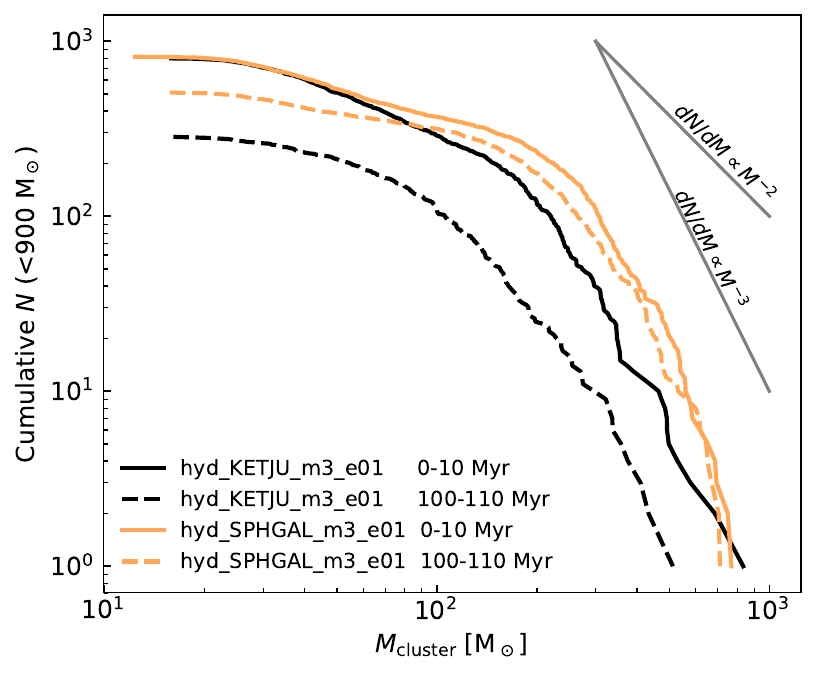}
\caption{The cumulative CMF of clusters in the age range of 0--10 Myr (solid) and 100--110 Myr (dashed) in the \texttt{hyd\_KETJU\_m3\_e01} (black) and \texttt{hyd\_SPHGAL\_m3\_e01} (orange) simulations. The CMFs are shown for clusters less massive than 900 \mdot{}, limited to include the same mass range in both simulations. The lines on the top right show power-law functions $dN/dM\propto M^{\alpha}$ with slopes $\alpha=-2$ and $\alpha=-3$.
\label{fig:CMF}}
\end{figure}

\begin{figure*}
\includegraphics[width=\textwidth]{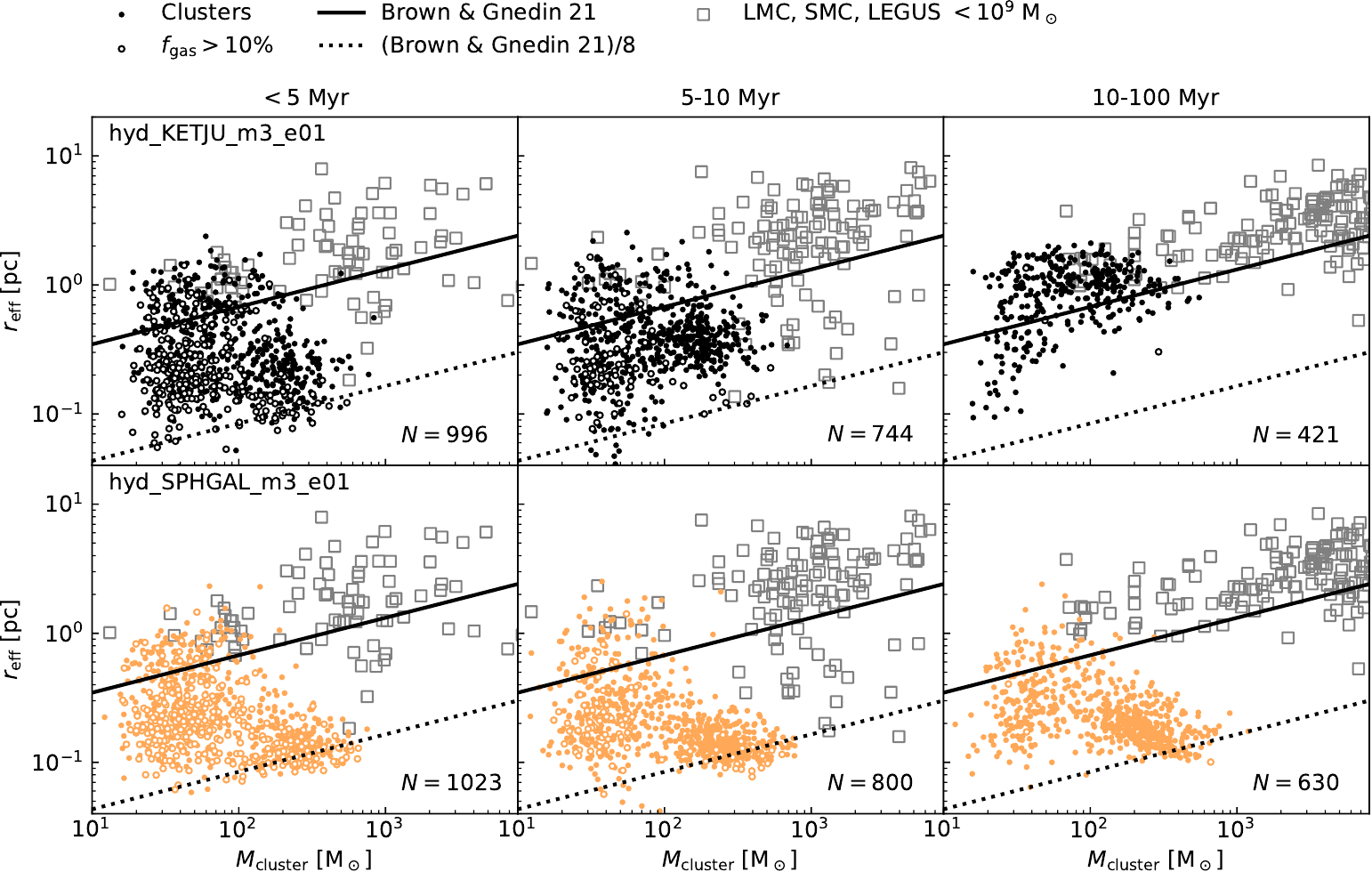}
\caption{The effective radius (2D) of bound star clusters (circles) in three age bins: 1--\mbox{5 Myr} (\textit{left}), 5--\mbox{10 Myr} (\textit{middle}) and 10--\mbox{100 Myr} (\textit{right}). The \textit{top row} shows the \texttt{hyd\_KETJU\_m3\_e01}-simulation and the \textit{bottom row} shows the \texttt{hyd\_SPHGAL\_m3\_e01}-simulation. Clusters that are still embedded, with at least 10\% of the mass within $r_\mathrm{eff}$ in gas, have been incidated with open circles. Open gray squares are observed results from \citet{2021MNRAS.508.5935B}, \citet{2003AJ....126.1836H} and \citet[][see text for details]{2021MNRAS.507.3312G}. The solid line is the best-fit relation from \citet{2021MNRAS.508.5935B} and the dotted line is the same relation scaled down by a factor of 8.
\label{fig:reffs}}
\end{figure*}

\subsection{Evolution of the star cluster population}

The cluster mass function (CMF) of all gravitationally bound star clusters is shown as a cumulative distribution in Fig. \ref{fig:CMF}. We show both young clusters (0--\mbox{10 Myr}) and evolved clusters (100--\mbox{110 Myr}). The clusters are stacked over 400 Myr of evolution (between 0--\mbox{400 Myr} for the 0--\mbox{10 Myr} age bin and 100--\mbox{500 Myr} for the 100--\mbox{110 Myr} age bin) in \mbox{10 Myr} steps so that the same clusters should appear once in both age-bins. We limit the mass-range to clusters less than $\sim900$ \mdot, which is the upper mass-limit of the clusters formed in the \ketju-simulations. The simulations without \ketju{} allow star clusters to grow on average to larger masses, thus we exclude three clusters more massive than $\sim900$ \mdot{} in the \texttt{hyd\_SPHGAL\_m3\_e01}-simulation from Fig. \ref{fig:CMF}.

The CMF in the \ketju-runs differ from the fully softened simulations both at young and at old cluster ages. Firstly, the \ketju-run has a shallower power-law shape at high cluster masses already at young cluster age. Observationally, the power-law slope of the CMF is often seen to be close to $-2$, thus the \ketju{} simulations seem to be in better agreement with the observations. However, the observed cluster samples are often more massive ($>10^{3-4}$ \mdot), have higher metallicity or are older, and are rarely complete at such low masses as considered here. We do not therefore attempt a direct comparison here. Other simulations of star cluster formation (e.g. \citealt{2022MNRAS.509.5938H, 2023MNRAS.522.2495G, 2024A&A...681A..28A}) have previously found that the shape of the CMF and mass of the the most massive cluster can also depend strongly on the stellar feedback and star formation prescriptions.

Second, the inefficient cluster disruption in the softened \sphgal-run results in very little evolution in the CMF over \mbox{100 Myr}, while the evolved CMF in the \ketju-run has significantly lower number of clusters in all mass-bins, highlighting again the efficient mass-loss. The total number of clusters formed in both runs is approximately the same. In total 63\% of the clusters by number that formed in the \ketju-run disrupt during the first 100 Myr of evolution, which is almost twice the corresponding value of 38\% of disrupted clusters in the \sphgal-run. 
 
\subsubsection{Expansion of young compact clusters}

In Fig. \ref{fig:reffs} we show the effective radii r$_\mathrm{eff}$ (2D in x--y plane) of clusters in three age bins of 1--5 Myr, 5--10 Myr and 10--100 Myr. The age ranges have been selected to enable direct comparison with observed star clusters in dwarf galaxies. We compare our cluster sizes to cluster data from the LEGUS survey \citep{2021MNRAS.508.5935B}, here limited to low-mass galaxies with $M_*<10^9$ \mdot, and clusters observed in the LMC and SMC from \citet{2003AJ....126.1836H} and \citet{2021MNRAS.507.3312G}. The radii used in \citet{2003AJ....126.1836H} are from \citet{1999AJ....117..238B}. These three samples were selected as they cover the full age and mass range considered here and provide an estimate of the half-mass or half-light radius, instead of e.g. a core-radius or a 90\% light radius as is often the case in Magellanic cloud cluster data. Fig. \ref{fig:reffs} shows also the best-fit relation from \citet{2021MNRAS.508.5935B} and this relation scaled down by a factor of 8. The latter illustrates the initial cluster sizes expected from N-body simulations that start with compact configurations and end on the observed size-mass relation \citep{2017A&A...597A..28B, 2024MNRAS.528.5119A}. 

The young clusters, especially more massive than 100 \mdot, have initial sizes that are similar or larger than the scaled \citet{2021MNRAS.508.5935B} relation in both simulations.
The \sphgal-clusters remain compact throughout their evolution. On the other hand, the increase in size of clusters simulated with \ketju{} can be seen already at very young ages. The youngest massive ($>100$ \mdot) \ketju-clusters have already started to show signs of expansion during the first 5 Myr. The 5--\mbox{10 Myr} clusters have already expanded by a factor of a few compared to their compact initial size, and the still older clusters that remain bound have effective radii up to 10 times larger than initially. The cluster sizes in the \ketju-simulation match very well with the observed sizes in their respective age bins, while the \sphgal-clusters run with softened gravity have too compact sizes at old age. In particular the more massive clusters are too compact in all age bins when no collisional dynamics is included. 

\begin{figure*}
\includegraphics[width=\textwidth]{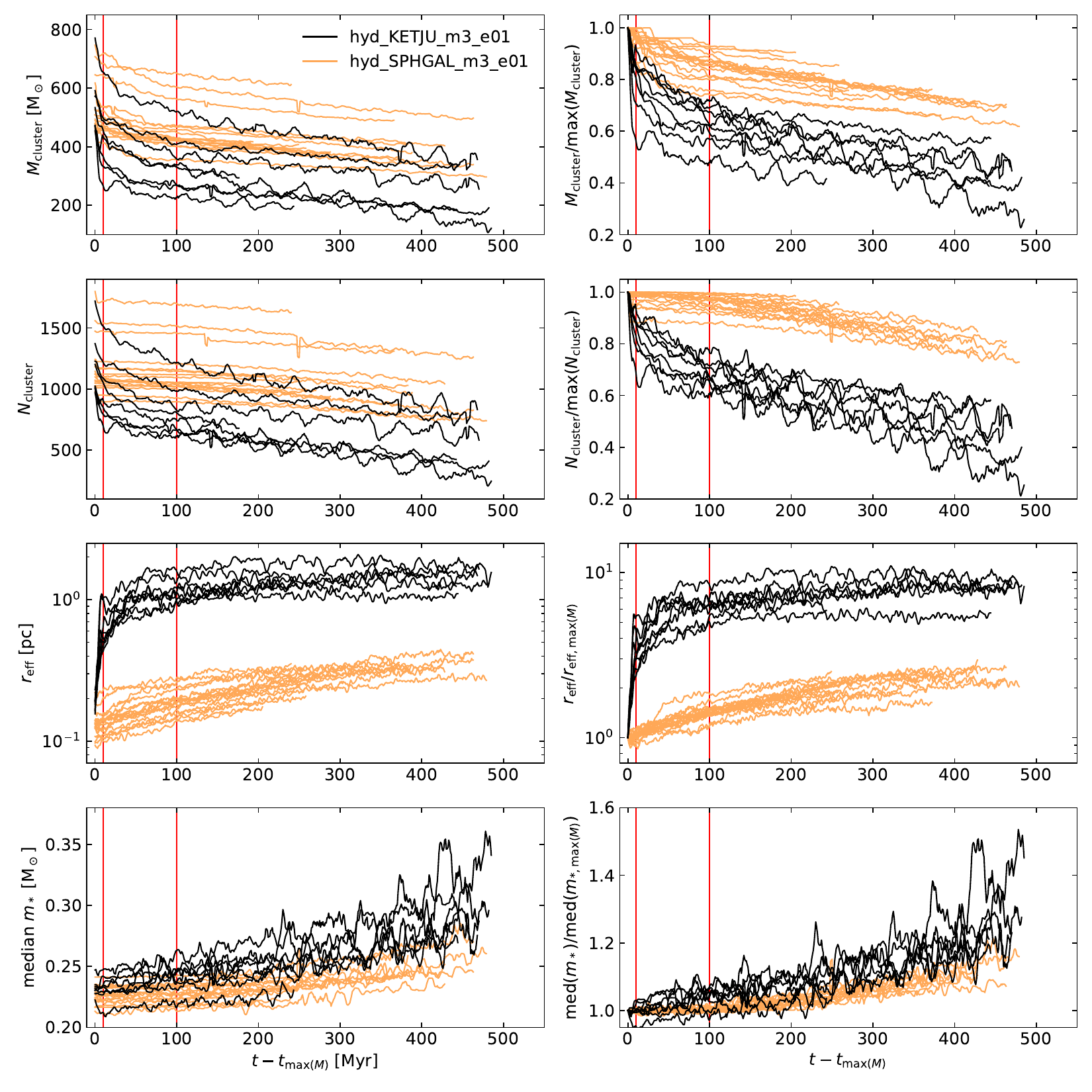}
\caption{From \textit{top} to \textit{bottom}: the bound mass, the number of bound stars, the effective radius (2D) and the median stellar mass in star clusters with maximum masses between 400--900 \mdot{} in the \texttt{hyd\_KETJU\_m3\_e01} (black) \texttt{hyd\_SPHGAL\_m3\_e01} (orange) simulations. The evolution is shown starting at the time of maximum cluster mass $t_\mathrm{max(\textit{M})}$. The \textit{left column} shows the absolute values and the \textit{right column} shows the values scaled by the equivalent value at $t_\mathrm{max(\textit{M})}$. The vertical lines indicate epochs of 10 and 100 Myr.
\label{fig:massloss}}
\end{figure*}

\subsection{Cluster evolution in a galactic context}\label{section:clusters}
\subsubsection{Mass-loss through stellar evolution and dynamics}

A more in-depth look into the evolution of individual clusters in the \ketju{} and \sphgal-simulations is provided in Fig. \ref{fig:massloss}. We have selected long-lived (at least 150 Myr) clusters with maximum masses in the range of 400--\mbox{900 \mdot{}} from both simulations (8 clusters in \texttt{hyd\_KETJU\_m3\_e01}, 17 clusters in \texttt{hyd\_SPHGAL\_m3\_e01}), and followed the evolution of their properties from birth until the final snapshot at \mbox{500 Myr}.  We show in the left panel of Fig. \ref{fig:massloss} (from top to bottom) the evolution of the bound mass of the cluster, the number of bound stars, the effective radius (2D, similar to Fig. \ref{fig:reffs}) and the median stellar mass. In the right panel of Fig. \ref{fig:massloss} these quantities are normalised to their initial value when the clusters reach maximum mass. 

The dynamical loss of bound stars in the \ketju-run is accompanied by rapid size-evolution already at very early cluster ages. This rapidly emerging effect is not seen in the \sphgal-run where the loss of stars is more gradual. Clusters in both simulations lose mass, especially through stellar evolution, but the softened simulation underestimates the dynamical mass-loss. It has been shown in N-body simulations that less extended clusters that have underfilled Roche lobes disrupt slower \citep{2005PASJ...57..155T, 2018ApJ...863..171S}. Thus the secondary effect of suppressed two-body interactions and the compact softening-supported size of 
the \sphgal-clusters is to lose mass at an even slower rate.

\begin{figure*}
\includegraphics[width=\textwidth]{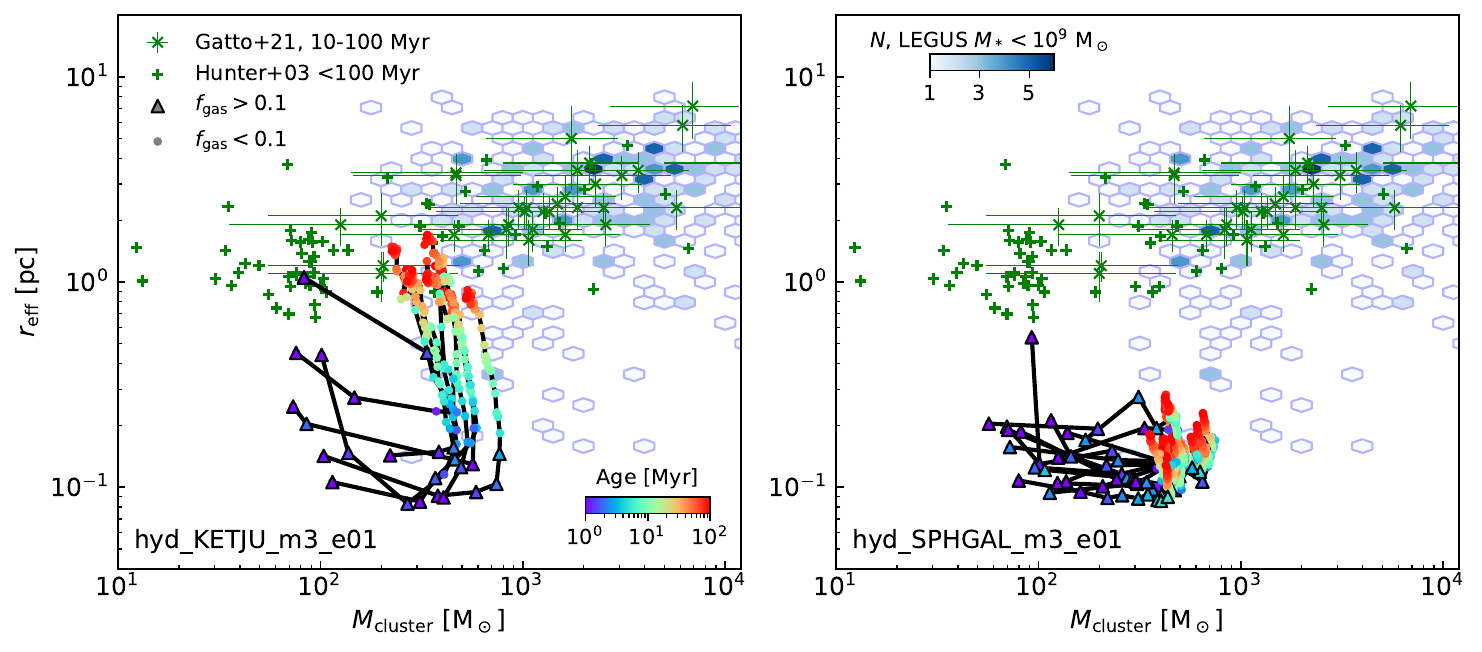}
\caption{The evolution of the effective radius (2D, measured in the x-y plane) and the bound stellar mass of star clusters with maximum masses between 400--900 \mdot{} in the \texttt{hyd\_KETJU\_m3\_e01} (\textit{left}) and \texttt{hyd\_SPHGAL\_m3\_e01} (\textit{right}) simulations. The data points along each evolution track have been coloured according to the cluster age, and the black-outlined triangles indicate epochs when the cluster is embedded in at least 10\% of its current mass in gas (within $r_\mathrm{eff}$). The observed data show young and intermediate age clusters ($<100$ Myr) in dwarf galaxies from the LEGUS dataset (blue hexbins, \citealt{2021MNRAS.508.5935B}) and the Magellanic clouds (green crosses from \citealt{2021MNRAS.507.3312G} and pluses from \citealt{2003AJ....126.1836H}).
\label{fig:reff_evol}}
\end{figure*}

The total lifetimes of star clusters can be estimated by fitting a mass-loss rate parametrised in \citet{2003MNRAS.340..227B} and \citet{2005A&A...441..117L} as
\begin{equation}
    \frac{dM_\mathrm{cluster}}{dt}=-\frac{\mathrm{M_\odot}}{t_0} \left(\frac{M_\mathrm{cluster}}{\mathrm{M_\odot}}\right)^{1-\gamma}
\end{equation} 
with time-scale parameter $t_0$ and $\gamma=0.7$. Here we set the initial mass of the fit $M_i$ at $t-t_\mathrm{max(\textit{M})}=100$ Myr, when the mass-loss has reached a gradual stage in the top panels of Fig. \ref{fig:massloss}. The best-fit total dissolution time $t_\mathrm{dis}=t_0(M_i/M_\odot)^{\gamma}$ is in the range of 0.8--2.3 Gyr for these selected clusters in the \ketju-simulation. The equivalent values for the \sphgal-run are between 2.0--4.2 Gyr. The total lifetimes of the clusters evolved with softened gravity in the tidal field of a dwarf galaxy are overestimated by at least factor of two at initial cluster mass of 400--900 \mdot. The discrepancy between the runs with and without \ketju{} increases with decreasing cluster mass, as the initial cluster mass is less clearly correlated with the total cluster lifetime. On the other hand, the smallest-mass clusters in the \ketju{}-simulation have always shorter estimated lifetimes compared to the more massive clusters.

The mass-segregation and escape of lower-mass stars is quantified in the median stellar mass in the bottom panel of Fig. \ref{fig:massloss}. The median stellar mass increases, in both simulations, which can only be the result of low-mass stars being removed from the stellar population. The median stellar masses of most of the \ketju-clusters evolve faster compared to the softened simulation. There is, however, some preferential removal of low-mass stars in the \sphgal-simulation as well. This indicates that the mass-loss process is partially but not entirely captured even when softened gravitational forces are used.

\subsubsection{Evolution in the mass-size plane}

In Fig. \ref{fig:reff_evol} we show the mass and size evolution of the individual clusters in the \texttt{hyd\_KETJU\_m3\_e01} and \texttt{hyd\_SPHGAL\_m3\_e01}-simulations from Fig. \ref{fig:massloss}, compared to the same observed mass-size data from Fig. \ref{fig:reffs}. The clusters are shown through 100 Myr of evolution starting from the first snapshot in which they were identified in, as opposed to the snapshot of maximum mass as in Fig. \ref{fig:massloss}. Here we include the early embedded phase, defined in this analysis as the period of time when the gas mass fraction within \mbox{1 pc} of the cluster centre of mass is larger than 10\%. We highlight these epochs to indicate when the clusters are embedded and when they later become fully exposed. Observed comparison data is the same as in Fig. \ref{fig:reffs}, here stacked across all clusters with ages less than 100 Myr located in low-mass galaxies ($M_*<10^9$ \mdot) from \citet{2021MNRAS.508.5935B}, \citet{2003AJ....126.1836H} and \citet{2021MNRAS.507.3312G}.

As was already shown in Fig. \ref{fig:massloss}, the \ketju-clusters evolve more rapidly toward the observed range of cluster sizes. The sizes first contract as the clusters grow in mass, but the radii begin increasing already when the clusters are still embedded. The \sphgal-clusters, on the other hand, reach maximum mass at high densities and remain in a narrow region of parameter space because neither the size nor the total mass evolve much as shown in Fig. \ref{fig:massloss}. 

The young clusters in both simulations are located in a region of parameter space that is not well populated by observed clusters. The observed comparison data shown in Figs. \ref{fig:reffs} and \ref{fig:reff_evol} extends down to young cluster ages but lacks the very youngest, embedded phase. Observations of the \textit{total} sizes of embedded clusters in the LMC \citep{2016ApJ...821...51R} span a range from $\sim0.3$ pc upward, with typical values of 1 pc. Low-mass embedded clusters in the Milky Way also exhibit small, sub-parsec sizes \citep{2007prpl.conf..361A}. The effective sizes of observed embedded clusters are thus expected to be smaller than the exposed clusters included in the comparison sample.  Our measured sizes in the embedded phase should thus be in agreement with the sizes of real embedded low-mass star clusters. 

The size-measurement technique we use is also not fully consistent with the observational method of measuring the cluster sizes. The sizes of the observed comparison clusters have been obtained through fitting a light-profile to the cluster emission. This makes the sizes sensitive to background estimation and difficult to measure in crowded environments. We instead only consider gravitationally bound stars when computing the sizes and masses of the clusters. The \ketju-clusters, in particular, have extended envelopes of unbound stars in the tidal tails that are excluded by the boundness criterion in the definition of our star clusters. A less strict clustering algorithm might thus result in even larger measured sizes at evolved cluster ages. The sizes measured through synthetic photometry might also differ from the measurements of direct particle data.

\begin{figure}
\includegraphics[width=\columnwidth]{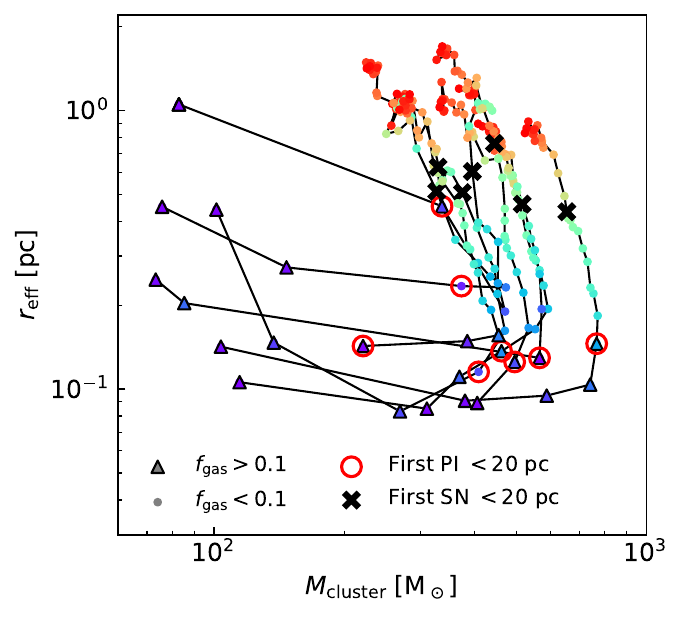}
\caption{The mass-radius evolution of clusters in the \texttt{hyd\_KETJU\_m3\_e01} simulation as shown in Fig. \ref{fig:reff_evol}. The time of the first phototionising star (red circles) and the first SN (black cross) within 20 pc of the cluster centre of mass. The clusters become gas free right after the first massive stars appear, while SNe occur only Myrs later.
\label{fig:expulsion}}
\end{figure}

\subsubsection{Gas expulsion}\label{section:gas_expulsion}

Exposed clusters in local galaxies are coincident with young stellar ages \citep{2014ApJ...795..156W, 2016MNRAS.460.2087H, 2018MNRAS.481.1016G, 2022MNRAS.512.1294H}, indicating that the embedded phase lasts only a few Myr. Fast gas expulsion is further supported by high-resolution simulations e.g. by \citet{2020MNRAS.499..748D} and \citet{2024MNRAS.527.6732F}. Based on the number of embedded data points per track in Fig. \ref{fig:reff_evol}, we see the bound clusters remain embedded for up to 5 Myr before full gas removal.

We further inspect the intersection of embedded and exposed clusters in the \texttt{hyd\_KETJU\_m3\_e01} \ketju-run in Fig. \ref{fig:expulsion} by showing the time of the birth of the first photoionising star and the first SN event within 20 pc of the cluster centre of mass superimposed on the mass-size evolution sequence from Fig. \ref{fig:reff_evol}. The embedded phase comes to an end when the average stellar age is a couple Myr\footnote{As opposed to the total stellar age spread or the total lifetime of the star cluster, which can be larger.} and often coincident with the ignition of the first photoionizing star. The first SNe occur only Myrs later, and do not contribute to the gas removal of the host cluster. We note that in the simulations of the present study, the shortest lifetimes of the SN-progenitor stars with masses up to 40 \mdot{} are $\sim 5$ Myr in the absence of stars in the pair-instability mass range. In a more intensely star-forming galaxy (e.g. \citealt{2024MNRAS.530..645L}), the onset of SNe could be hastened through the formation of more massive star clusters that are able to host more massive stars with lifetimes as short as a couple Myrs.

The small sizes and the gas expulsion induced by photoionisation in our simulated young clusters therefore seem to be in agreement with observed star clusters, and a picture emerges wherein the clusters evolve rapidly toward lower central densities during and after gas removal. The driver of gas removal is photoionising radiation, well before SNe in the vicinity of the clusters come into play. This result is in agreement with previous simulations that have shown that a sufficiently luminous source can significantly disrupt the gas cloud before the SNe occur \citep{2013MNRAS.430..234D, 2016MNRAS.463.3129G, 2018ApJ...859...68K, 2019MNRAS.489.1880H, 2020MNRAS.492.4858H, 2020MNRAS.497.3830F, 2022MNRAS.515.4929G}, and our results extend these previous works with a model that includes individually resolved radiation sources in a full galactic context. The present results are also in qualitative agreement with previous analysis of the simulations in the \griffin{}-project on starburst systems (\citealt{2024MNRAS.534..215F}) where we have shown that even the most massive cold star-forming ISM-clouds ($>10^5$ \mdot) can be destroyed already after the first SN. The pre-processing by early stellar feedback is thus a crucial driver of evolution of clustered star-forming regions across environments, though full gas removal by photoionisation might be only possible in relatively low-mass clusters. The drastic difference in the size-evolution of the collisional and softened clusters in Fig. \ref{fig:reff_evol} however highlights the fact that the full impact of gas dispersal and removal on the evolution of star clusters can only be addressed in hydrodynamical simulations that account for their collisional nature. The internal evolution of more massive clusters with a prolonged embedded phase will be addressed in a future study.

\section{Summary and Conclusions}\label{section:conclusions}

We present new hydrodynamical+collisional N-body simulations of star formation in low-metallicity dwarf galaxies, concentrating on the formation and evolution of gravitationally bound star clusters. The simulations are run with the modern SPH code \sphgal{} now also including regularised small-scale gravitational dynamics using the \ketju{} integrator. We have updated the interface between \sphgal{} and \ketju{} which enables us to place regularised integration regions around all selected stars above a mass threshold and their remnants, as outlined in Table \ref{table:ics}. This enables collisional dynamics within star clusters in the vicinity of massive stars to be resolved accurately without the need for gravitational softening. 

We first benchmark the updated \sphgal+\ketju{} method with the direct N-body code \bifrost{} by running idealised star clusters in isolation without stellar evolution. \sphgal+\ketju{} captures the mass-segregation and two/few-body interactions within the central regions of the clusters, thus producing similar numbers of unbound stars (to within a few tens of per cent) as the direct N-body code. These stars escape the cluster as runaway or walkaway stars even without an external tidal field. Because the majority of interactions between lower mass stars in the outskirts of clusters are still softened, the evolution toward energy equipartition is slowed down. This results in a somewhat slower expansion of the clusters in the \sphgal+\ketju-simulations compared to \bifrost. When the number of \ketju-regions is restricted to only stars more massive than 8 \mdot{}, only a few per cent of the gravitational interactions are regularised. Even then, the clusters still expand significantly and produce runaway stars, owing to mass-segregation and strong few-body interactions in the central regions of the clusters.

We then analyse the long-term evolution of low-metallicity dwarf galaxies integrated with the \sphgal+\ketju{} code, in comparison to simulations that adopt gravitational softening everywhere. The global properties, such as the total new stellar mass, the outflow rate and mass-loading factors are quite similar, with differences in the cumulative properties only of the order of a few tens of per cent. SNe are less clustered in the \ketju-run with regularisation due to cluster dissolution, which leads to less efficient recycling of wind and SN-material in new stars. Most of the wind and SN-material still escapes the galaxy in metal-enriched outflows. Photoionisation regulates the gas densities, thus the lowered SN-clustering has a little impact on the global star-forming environment.

The major impact of adding accurate gravitational interactions is seen in the formation and evolution of star clusters in the galactic tidal field. The average values of $\Gamma$ for clusters in the age interval of 1--10 Myr are lowered from $\sim 40\%$--$50\%$ in the softened simulation to $\sim30\%$--$35\%$ in the \ketju-run due to the rapid loss of stars from the clusters. Evolved clusters in the range 10--100 Myr contain only $\sim13\%$ of all stellar mass in the system in the \ketju-run. This is in agreement with observations where $\Gamma$ at older cluster age can be used to probe cluster dissolution when cluster mass-loss cannot be directly followed. Approximately 63\% of the clusters by number in the \ketju-simulation dissolve completely during the first $\sim100$ Myr of evolution, compared to 38\% in the run where softening suppresses cluster disruption.

We also took a more detailed look at the evolution of a limited sample of individual long-lived clusters formed in the hydrodynamical simulations, with initial masses of a few hundred solar masses. The clusters in both simulations suffer mass-loss through stellar winds and SNe. The \ketju-runs have a large fraction (up to $40\%$ by number) of initially bound stars that become unbound during the first 10--100 Myr. The $r_\mathrm{eff}$ of selected clusters in the initial mass-range of 400--900 \mdot{} are shown to increase by a factor of 2--6 during the first 10 Myr of evolution, and by a factor of 5--9 during the first \mbox{100 Myr}. Also the fully softened clusters expand somewhat, by a factor of 2--3, and they lose bound stars (up to $\sim10\%$) due to the interactions with the galactic tidal field. This is in contrast with the pure N-body simulations of isolated star clusters where the structure of the fully softened clusters did not evolve nearly at all. The mass-loss in the softened clusters is, however, too slow by at least by a factor of two.

As the \ketju-clusters evolve in the mass-size plane, they end up in the observed range of parameter space that is occupied by exposed relatively young star clusters in low-mass galaxies. We showed that the sizes first contract while clusters are still forming stars, but the sizes can start increasing already when the clusters are still embedded. The sizes then increase rapidly once the first photoionising stars are ignited and the clusters become fully exposed. The gas expulsion is governed by the pre-SN stellar feedback while SNe occur only Myrs after the gas has already been removed. 

In this study we discuss clusters less massive than 1000 \mdot. The star formation time scale is longer \citep{2019MNRAS.489.1880H, 2020ApJ...891....2L} and the gas retention may be more efficient in more massive clusters as indicated in our previous work on star-by-star starburst environments in \citet{2024MNRAS.530..645L}. In more massive clusters, where the gas can be retained in the cluster centre for a longer period of time, stellar winds and SN feedback may become more important for chemical enrichment and gas removal. Investigation of more massive clusters will be left for a future study, where we will apply the new hydrodynamical+N-body method in a low-metallicity starburst. An implementation for stellar interactions and mergers would enable us to study the collisional growth of very massive stars in dense star clusters \citep{2004Natur.428..724P}. Such dense cluster-forming regions may also have given rise to globular clusters in the early Universe \citep{2024Natur.632..513A}, which may have been important sources of ionising radiation \citep{2020MNRAS.492.4858H}. The rapid (Myr-scale) collisional growth of massive stars in these dense clusters may provide a channel for enhanced chemical enrichment \citep{2018MNRAS.478.2461G} and black hole formation in the mass range of \mbox{$\sim10^3$ \mdot{}}--\mbox{$10^4$ \mdot{}} \citep{2024Sci...384.1488F, 2024MNRAS.531.3770R}. With this simulation methodology we may begin to address the origin of the chemical enhancements \citep{2023A&A...677A..88B} and the seeds of the supermassive black holes \citep{2024NatAs...8..126B} recently uncovered with JWST at unexpectedly high redshifts.

\section*{Acknowledgements}

NL thanks Deidre Hunter and Bruce Elmegreen for providing the data reported in \citet{2003AJ....126.1836H}. NL and TN gratefully acknowledge the Gauss Centre for Supercomputing e.V. (www.gauss-centre.eu) for funding this project by providing computing time on the GCS Supercomputer SUPERMUC-NG at Leibniz Supercomputing Centre (LRZ, www.lrz.de) under project numbers pn49qi and pn72bu. TN acknowledges support from the Deutsche Forschungsgemeinschaft (DFG, German Research Foundation) under Germany's Excellence Strategy - EXC-2094 - 390783311 from the DFG Cluster of Excellence "ORIGINS". PHJ acknowledges the support by the European Research Council via ERC Consolidator Grant KETJU (no. 818930) and the support of the Academy of Finland grant 339127.
The computations were carried out at SuperMUC-NG hosted by the LRZ and the COBRA and FREYA clusters hosted by The Max Planck Computing and Data Facility (MPCDF) in Garching, Germany.

This research made use of \textsc{python} packages \textsc{scipy} \citep{2020SciPy-NMeth}, \textsc{numpy} \citep{2020NumPy-Array}, \textsc{matplotlib} \citep{Hunter:2007}, \textsc{pygad} \citep{2020MNRAS.496..152R}, \textsc{h5py} \citep{collette_python_hdf5_2014} and \textsc{astropy}\footnote{http://www.astropy.org} \citep{2022ApJ...935..167A}.

\section*{Data Availability}

The data will be made available on reasonable request to the corresponding author.



\bibliographystyle{mnras}



\appendix\label{section:appendix1}
\section{Central 5 per cent Lagrangian radii of individual clusters}

\begin{figure*}
\includegraphics[width=0.9\textwidth]{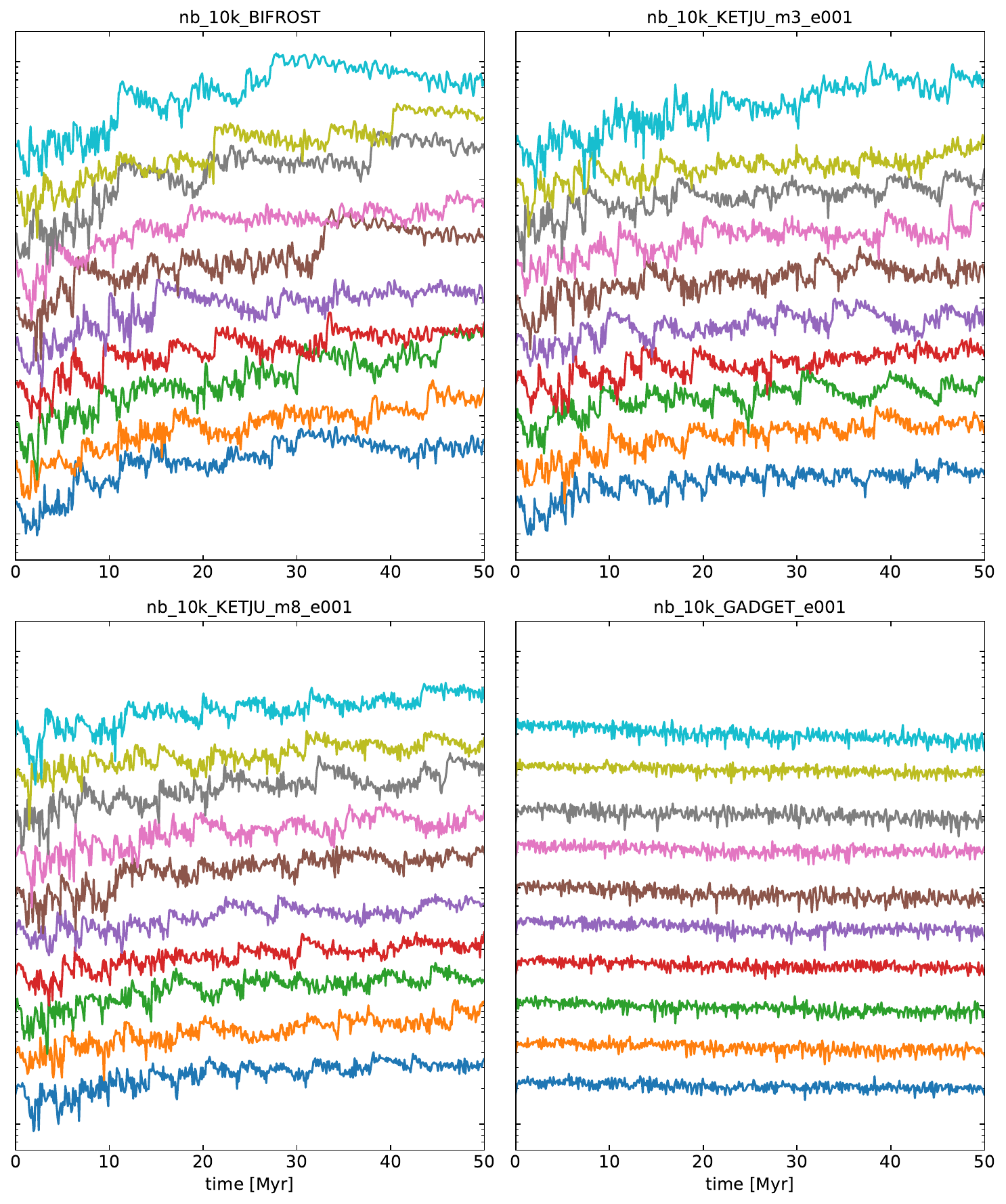}
\caption{The evolution of the 5\% Lagrangian radii of the ten individual randomly generated clusters run in Section \ref{section:nbody_results}. The lines from top to bottom in each panel show the simulations started from the same initial condition. The lines have been shifted to ease their differentiation, therefore the y-axis is arbitrary. The simulation samples from left to right, top to bottom are \texttt{nb\_10k\_BIFROST}, \texttt{nb\_10k\_KETJU\_m3\_e001}, \texttt{nb\_10k\_KETJU\_m8\_e001} and \texttt{nb\_10k\_GADGET\_e001}.
\label{fig:appendix1}}
\end{figure*}

Fig. \ref{fig:appendix1} shows the inner 5\% Lagrangian radii in each individual simulation of the code comparison sample discussed in Section \ref{section:lagrangian}. The mean standard deviation of the simulations shown in each panel of Fig. \ref{fig:appendix1} are shown in Fig. \ref{fig:nbody_lagrangian}. The cycles of contraction and rapid expansion are most clearly seen in the \texttt{BIFROST} runs. The \texttt{KETJU}-simulations do also show these cycles, albeit often with a smaller amplitude. The inner regions of the \texttt{GADGET}-simulations show practically no evolution.


\bsp	
\label{lastpage}
\end{document}